\documentclass[fleqn,10pt]{wlscirep}
\usepackage[utf8]{inputenc}
\usepackage[T1]{fontenc}
\usepackage{graphicx}
\usepackage{epstopdf, epsfig}
\usepackage{amsmath}
\usepackage{mathtools}
\usepackage{cancel} 
\usepackage{hyperref}
\hypersetup{
    colorlinks=true,
    linkcolor=blue,
    citecolor=blue
    }
\usepackage{xcolor}
\usepackage{float}
\usepackage{caption}
\usepackage{subcaption}
\usepackage{dsfont}

\newcommand{\corr}[1]{\textcolor{black}{#1}}

\title{The coherent structure of the energy cascade in isotropic turbulence}

\author[1,+]{Danah Park}
\author[2,3,*,+]{Adri\'an Lozano-Dur\'an}
\affil[1]{Mechanical Engineering Department, Stanford University, Stanford, CA 94305, USA}
\affil[2]{Department of Aeronautics and Astronautics, Massachusetts Institute of Technology, Cambridge, MA 02139}
\affil[3]{Graduate Aerospace Laboratories, California Institute of Technology, Pasadena, CA 91125, USA}

\affil[*]{adrianld@mit.edu}
\affil[+]{these authors contributed equally to this work}

\begin{abstract}
The energy cascade, i.e. the transfer of kinetic energy from
large-scale to small-scale flow motions, has been the cornerstone of
turbulence theories and models since the 1940s. However, understanding
the spatial organization of the energy transfer has remained elusive.
In this work, we answer the question: What are the characteristic flow
patterns surrounding regions of intense energy transfer?  To that end,
we utilize numerical data of isotropic turbulence to investigate the
three-dimensional spatial structure of the energy cascade in the
inertial range. Our findings indicate that forward energy-transfer
events are predominantly confined in the high strain-rate region
created between two distinct zones of elevated enstrophy.  On average,
these zones manifest in the form of two hairpin-like shapes with
opposing orientations. The mean velocity field associated with the
energy transfer exhibits a saddle point topology when observed in the
frame of reference local to the event.  The analysis also shows that
the primary driving mechanism for the cascade involves strain-rate
self-amplification, which is responsible for 85\% of the energy
transfer, whereas vortex stretching accounts for less than 15\%.
\end{abstract}
\begin{document}

\flushbottom
\maketitle

\section*{Introduction}

Turbulence exhibits a wide range of flow scales, whose non-linear
interactions still challenge our intellectual ability to understand
even the simplest flows.  These interactions are responsible for the
cascading of kinetic energy from large eddies to the smallest eddies,
where the energy is finally dissipated~\cite{Richardson1922,
  Obukhov1941, Kolmogorov1941}. Given the ubiquity of turbulence, a
deeper understanding of the energy transfer among flow scales would
enable significant progress to be made across various fields ranging
from combustion~\cite{Veynante2002},
meteorology~\cite{Bodenschatz2015}, and
astrophysics~\cite{Young2017} to engineering applications of
external aerodynamics and hydrodynamics~\cite{Sirovich1997,
  Hof2010, Marusic2010, Kuhnen2018}.

The phenomenological description of the turbulent cascade in terms of
interactions among eddies at different scales was first proposed by
Richardson~\cite{Richardson1920} and later by
Obukhov~\cite{Obukhov1941}. Since then, substantial efforts have been
directed toward the characterization of inter-scale kinetic energy
transfer.  The statistical description of the transfer of energy from
large to small scales was introduced in the classical paper by
Kolmogorov~\cite{Kolmogorov1941}. Since then, a large body of research
has been devoted to addressing two outstanding questions about the
energy cascade: 1) Is the transfer of kinetic energy from large to
small scales local in scale? and 2) What are the physical mechanisms
driving the transfer of energy among scales? Here, we pose a third
question: 3) What are the characteristic flow patterns surrounding
regions of intense energy transfer?  The three questions above are
interconnected, and we anticipate that addressing question 3) will
also shed light on question 2).

There is general agreement from the community that the answer to the
first question is yes.  The net transfer of energy from large to small
scales is mainly accomplished by interactions among flow motions of
similar size.  The conclusion is supported by evidence from diverse
approaches, such as scaling analysis~\cite{ Zhou1993a, Zhou1993b,
  Eyink1995, Aoyama2005, Eyink2005, Mininni2006, Mininni2008,
  Aluie2009, Eyink2009}, triadic interactions in Fourier
space~\cite{Domaradzki1990, Domaradzki2009}, time
correlations~\cite{Cardesa2015, Cardesa2017}, and
information-theoretic causality~\cite{lozano2023, Martinez2024}, to
name a few.
%

The degree of consensus is lower regarding the second question.
Several theories have been proposed to describe the physical
mechanism(s) that enable the transfer of energy from larger to smaller
scales throughout the inertial range.  One of the first mechanisms
proposed is based on the concept of vortex stretching.  In this
scenario, vorticity is stretched by the strain-rate either at the same
scale or at a larger scale~\cite{Taylor1937, Taylor1938, Tennekes1972,
  Davidson2008, Hamlington2008, Leung2012, Lozano2016, Doan2018}. Some
authors have further proposed that the mechanistic details of the
process can be explained by successive reconnections of anti-parallel
vortex tubes~\cite{Melander1988, Hussain2011, Goto2017, Yao2020},
vortex reconnection of two long, straight anti-parallel vortex tubes
with localized bumps~\cite{Kerr2013}, and the presence of helical
instabilities in vortex rings~\cite{Brenner2016, Mckeown2018}. These
viewpoints, while not dynamically equivalent, are still compatible
with the vortex-stretching driven energy cascade.
%
%

%
%
The main competing theory to the vortex stretching mechanism is the
self-amplification of the rate-of-strain either by same-scale
interactions or by the amplification from larger scale
strain-rate~\cite{Tsinober2009, Paul2017, Sagaut2008,
  Carbone2020, Vela2021}. In both scenarios, the strain-rate
self-amplification stands as the key contributor to the transfer of
energy among scales, whereas vortex stretching is merely the effect
(rather than the cause) of the energy cascade.
%
%
Given the kinematic relationship between vortex stretching and
strain-rate self-amplification~\cite{Betchov1956, Capocci2023}, some
authors have argued that both mechanisms play a relevant role in the
dynamics of the energy cascade as one cannot occur without the
other~\cite{Johnson2020, Johnson2021}.

The studies above have helped advance our understanding of the physics
of the energy cascade; however, less is known about flow patterns
associated with the cascading process.  With the advent of novel flow
identification techniques~\cite{Delalamo2006, Lozano2012,
  Lozano2014, Dong2017, Dong2020}, the three-dimensional
characterization of turbulent structures is now achievable to complete
the picture.  In this work, we shed light on the characteristic flow
patterns associated with the energy cascade by investigating the
spatial three-dimensional structure of the flow conditioned to intense
energy transfer events.


\section*{Methods}
\label{sec:method}

\subsection*{Numerical dataset and filtering procedure}

We use direct numerical simulations of isotropic
turbulence~\cite{Cardesa2015}. The numerical setup corresponds to
isotropically forced turbulence within a triply periodic domain at
$\mathrm{Re}_\lambda = 146, 236$, and $384$, where
$\mathrm{Re}_\lambda$ is the Reynolds number based on the Taylor
microscale.  The simulations are labeled as HIT1 (for
$\mathrm{Re}_\lambda = 146$), HIT2 (for $\mathrm{Re}_\lambda = 236$),
and HIT3 (for $\mathrm{Re}_\lambda = 384$).  Table~\ref{table:setup}
contains the numerical details of the simulations in terms of grid
resolution, domain size, and number of snapshots.  The reader is
referred to Ref.~\cite{Cardesa2015} for additional information about
the numerical setup.
\begin{table}
\centering
\begin{tabular}{ c c c c c c c c }
    Case  & $\mathrm{Re}_\lambda$  & $N_x \times N_y \times N_z$   & $\left(L_x \times L_y \times 
    L_z\right)/\eta$   & $L_o/\eta $   & $N_t$ & $\Delta/\eta$ & $\Delta t/T_o$  \\
    HIT1 & $146$ & $256^3$ & $506^3$ & $425$ & 30 & 2 & 5.5 \\
    HIT2 & $236$ & $512^3$ & $1011^3$ & $876$ & 48 & 2 & 0.33  \\
    HIT3 & $384$ & $1024^3$ & $2022^3$ & $1813$ & 30 & 2 & 0.13 \\
\end{tabular}
\caption{\label{table:setup} Summary of the main parameters of
    the simulations. $\mathrm{Re}_\lambda$ is the Reynolds number
    based on the Taylor microscale. $N_i$ and $L_i$ are the number of
    spatial Fourier modes and the domain size in the directions $i=x,
    y, z$. The Kolmogorov and integral length scales are
    $\eta=\left(\nu^3 / \bar{\varepsilon}\right)^{1 / 4}$ and
    $L_o=\bar{K}^{3 / 2} / \bar{\varepsilon}$, respectively, where
    $\bar{K}$ and $\bar{\epsilon}$ are the space-time averaged
    turbulent kinetic energy and dissipation. The times are normalized
    by $T_o=\bar{K} / \bar{\varepsilon}$ and $N_t$ is the number of
    instantaneous flow fields used to collect
    statistics; $\Delta/\eta$ is the grid resolution and $\Delta
    t/T_o$ is the time interval between instantaneous flow fields used
    for the analysis.}
\end{table}

\begin{figure}
    \centering
    \subfloat[]{\includegraphics[width=.48\linewidth]{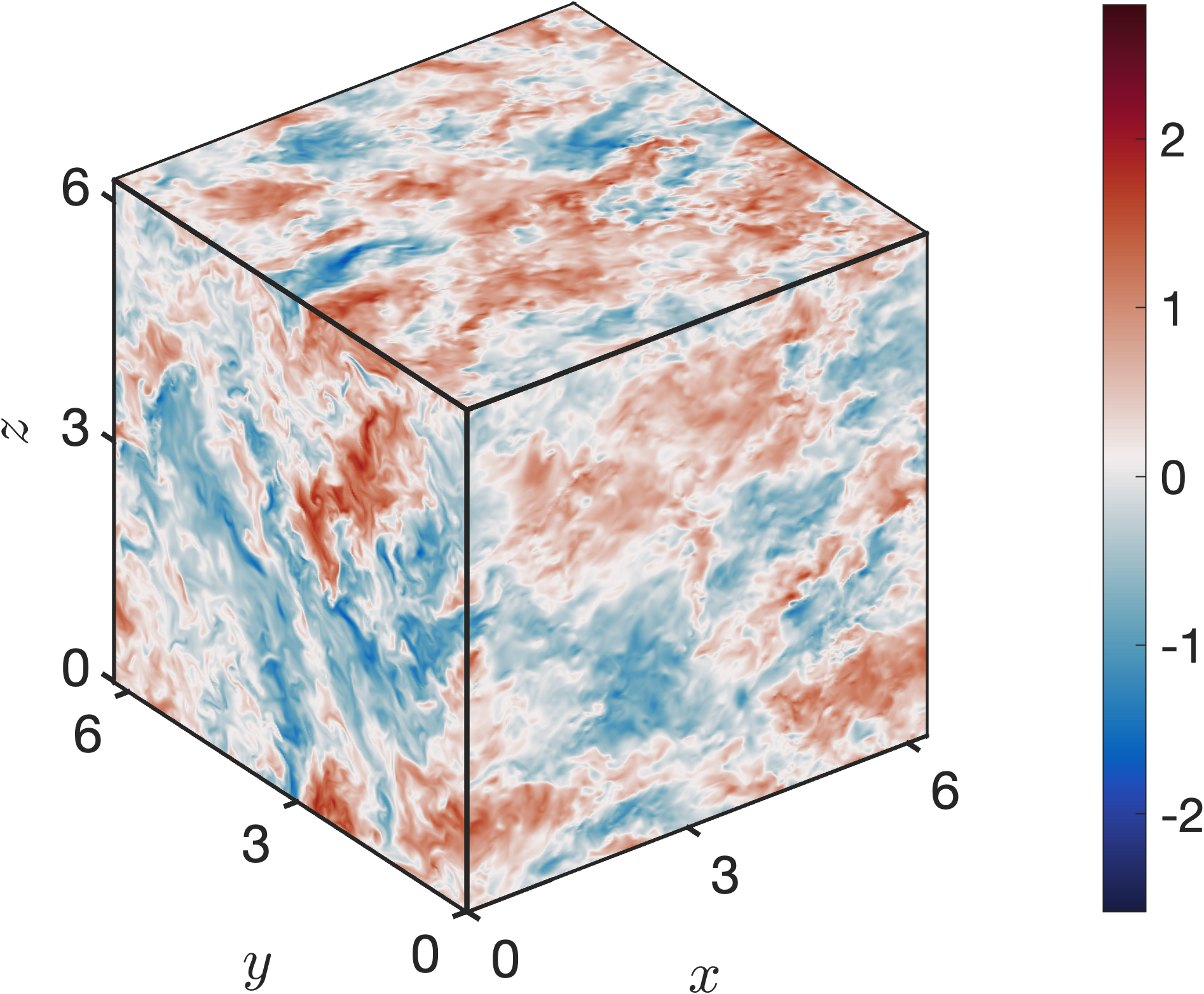}}
    \hspace{0.2cm}
    \subfloat[]{\includegraphics[width=.48\linewidth]{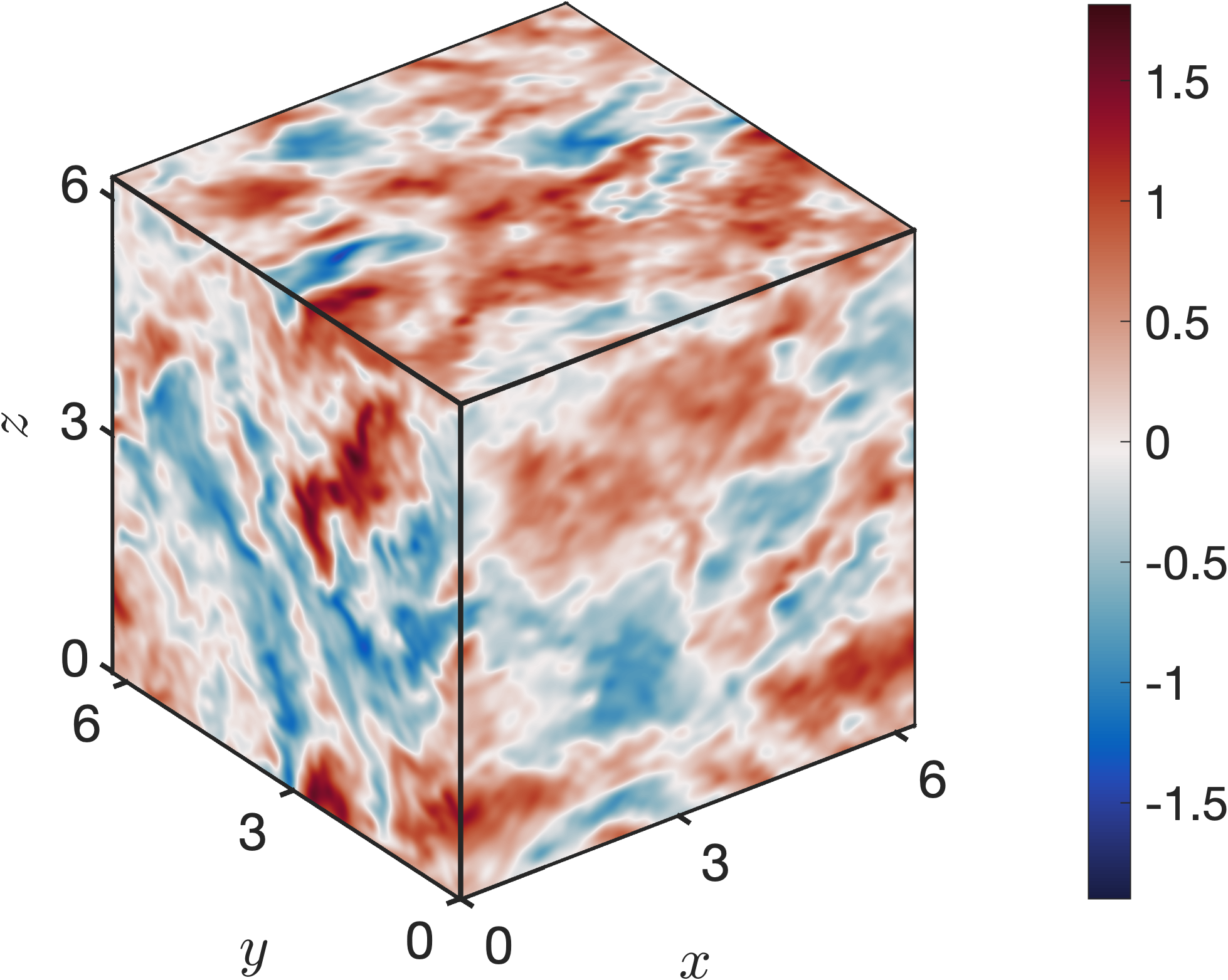}}
    \caption{Instantaneous velocity field for (a) $u_1$ (unfiltered)
      and (b) $\widetilde{u}_1$ (filtered) for $r = 30\eta$. The
      velocity is normalized by $\sqrt{\langle u_i u_i \rangle}$ in
      panel (a) and by $\sqrt{\langle \tilde{u}_i \tilde{u}_i}
      \rangle$ in panel (b), where $\langle \cdot \rangle $ denotes
      average over homogeneous directions and time.  The visuals in
      this figure, along with those in Figures 3, 5, 7, and 8, were
      generated using Python
      3.12~\cite{Python312}. \label{fig:velocity}}
\end{figure}

\begin{figure}
    \centering
    \subfloat[]{\includegraphics[width=.48\linewidth]{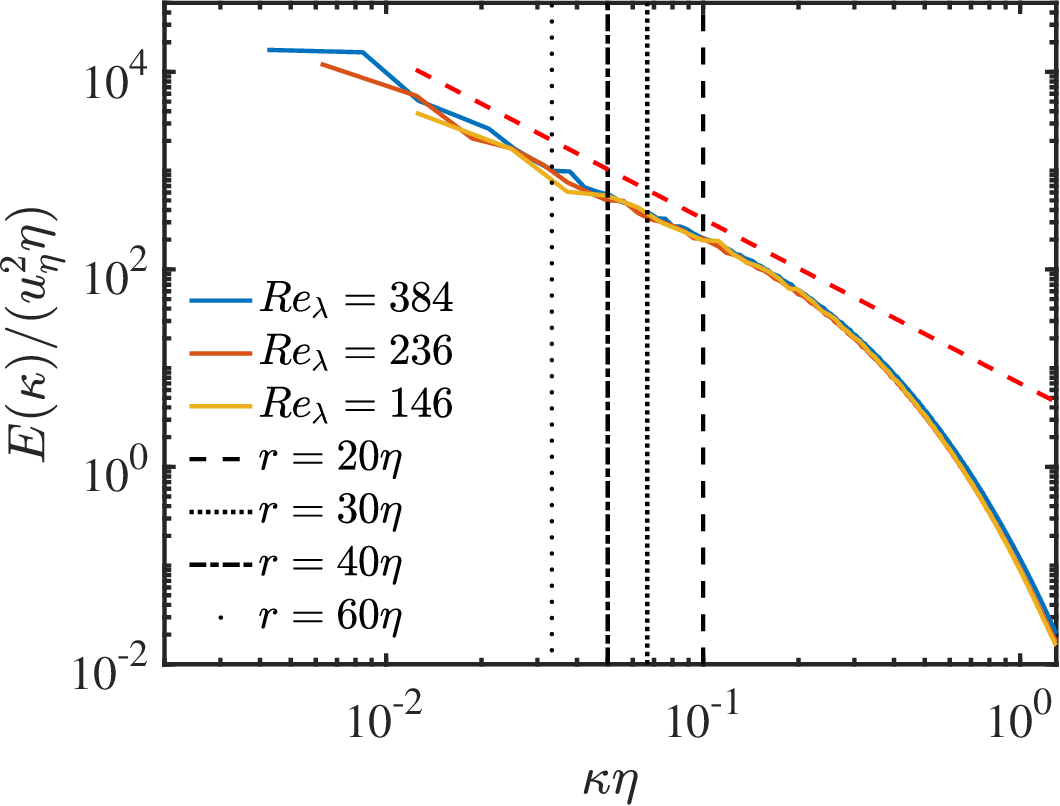}}
    \hspace{0.2cm}
    \subfloat[]{\includegraphics[width=.48\linewidth]{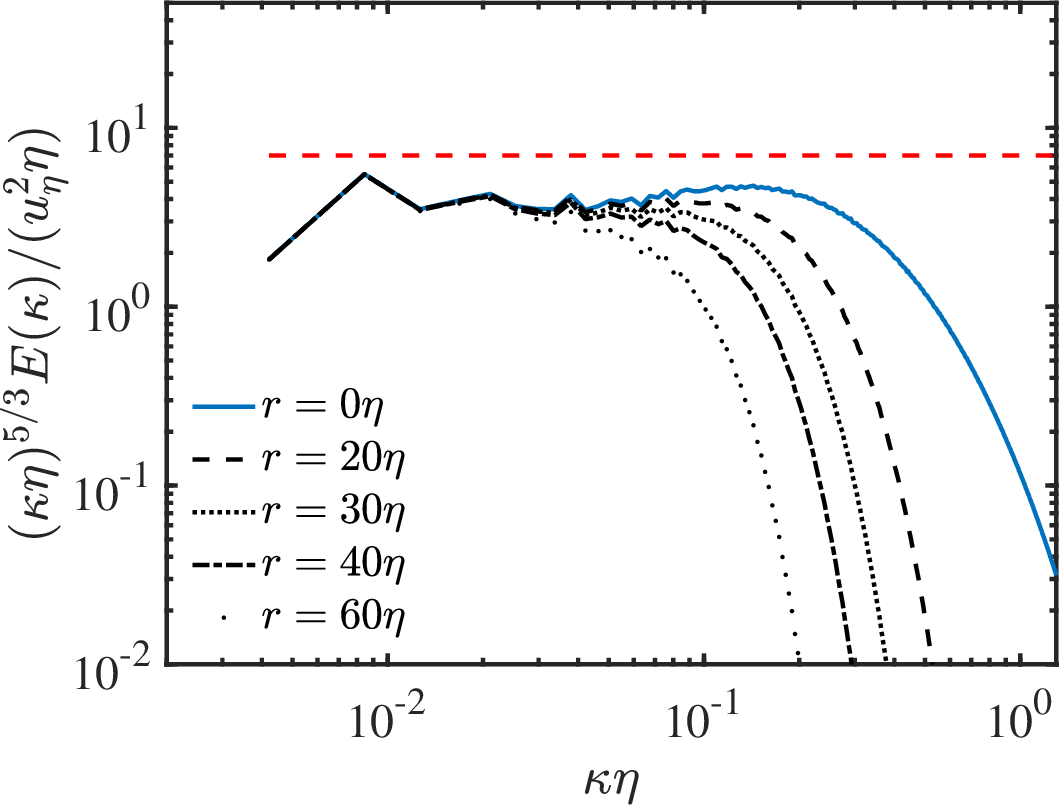}}
        \caption{The kinetic energy spectra $E(\kappa)$ as a function
          of the wavenumber $\kappa$. The spectra and wavenumbers are
          normalized with the Kolmogorov length-scale $\eta$ and the
          Kolmogorov velocity-scale $u_\eta$. (a) $E(\kappa)$ for the
          unfiltered velocity. The vertical lines correspond to the
          different filter widths.  (b) Compensated kinetic energy
          spectra $\kappa^{5/3}E(\kappa)$ for different filtered
          widths and for case HIT3.  The red dashed line in both
          panels is $E(k) \sim k^{-5/3}$.  \label{fig:spectra}}
\end{figure}

To investigate the structure of the energy transfer, the velocity
field is decomposed into large and small scales using a low-pass
Gaussian filter such that $u_i(\mathbf{x},t) =
\widetilde{u}_i(\mathbf{x},t) + u'_i(\mathbf{x},t)$ for $i=1,2,3$,
where ${u}_i$ represents the $i$-th component of the instantaneous
velocity, $\widetilde{u}_i$ is the filtered velocity, and $u'_i$
corresponds to the remaining fluctuating velocity component. The
spatial coordinate is $\mathbf{x} \equiv [x_1,x_2,x_3]$. Occasionally,
we use the notation $[u_1,u_2,u_3] \equiv [u,v,w]$ for velocities and
$[x_1,x_2,x_3]=[x,y,z]$ for space. The filter is isotropic and given
in Fourier space by the Gaussian kernel $\hat{G}(\mathbf{\kappa}) =
\exp[-(r\kappa)^2/24]$, where $\kappa=\mathbf{|\kappa|}$ is the
magnitude of the wavenumber vector ($\mathbf{\kappa}$), and $r$ is the
filter width.  Four filter widths are investigated: $r/\eta = 20, 30,
40,$ and $60$, where $\eta= \left(\nu^3 / \bar{\varepsilon}\right)^{1
  / 4}$ is the Kolmogorov length-scale, $\nu$ is the kinematic
viscosity, and $\bar{\epsilon}$ is the space-time averaged dissipation
of turbulent kinetic energy.  Figure~\ref{fig:velocity} compares one
instant of the unfiltered velocity and the filtered velocity for $r =
30\eta$.  The values of $r$ were chosen to span across the inertial
range of the turbulence cascade~\cite{Cardesa2015}. The kinetic energy
spectra $E(\kappa)$ for the three Reynolds numbers are shown in
figure~\ref{fig:spectra}(a). As expected, the viscous and inertial
range of $E(\kappa)$ collapse in Kolmogorov units, defined by $\eta$
and the Kolmogorov velocity-scale $u_\eta = (\nu
\bar{\epsilon})^{1/4}$.  The compensated kinetic energy spectra
$\kappa^{5/3}E(\kappa)$ for HIT3 is included in
figure~\ref{fig:spectra}(b) for the unfiltered case and for
$r/\eta=20, 30, 40$ and $60$. The latter shows that the filtered
widths lie within the inertial range of the moderate Reynolds numbers
considered.

\corr{It is worth noting that other filtering approaches are also
  possible. Among them, we can mention the box filter (i.e., filter
  width fully localized in physical space) and the Fourier cut-off
  filter (i.e., filter width fully localized in wavenumber
  space). Although not shown, we compared the averaged energy transfer
  patterns using box filters and Fourier cut-off filters and obtained
  qualitatively similar results to those presented below. In this
  work, we focus our analysis on the Gaussian filter, as it offers a
  compromise between the box and Fourier filters while allowing for
  the decomposition of energy transfer mechanisms later discussed in
  Eq.~(\ref{eq:GK_terms}).}

\subsection*{Identification of intense energy transfer events}
\label{sec:structures}

The kinetic energy equation for the filtered velocity field is
\begin{equation}
\frac{D}{Dt} \left( \frac{1}{2} \widetilde{u}_{i} \widetilde{u}_{i} \right)=-\frac{\partial}{\partial x_{j}}\left(\widetilde{u_{j}} \widetilde{p}+\widetilde{u}_{i} \tau_{i j}-2 \nu \widetilde{u}_{i} \widetilde{S}_{i j}\right)-2 \nu \widetilde{S}_{i j} \widetilde{S}_{i j}-\Pi+\widetilde{u}_{i}{\widetilde{f_{i}}},
\label{eq:filteredKE}
\end{equation}
where repeated indices imply summation, ${D} / {Dt} = {\partial} /
{\partial t}+\widetilde{u}_{j} {\partial} / {\partial x_{j}}$ is the
material derivative, $\widetilde{S}_{i j}=\left(\partial
\widetilde{u}_{i} / \partial x_{j}+\partial \widetilde{u}_{j} /
\partial x_{i}\right) / 2$ is the strain-rate tensor for the filtered
velocity, and $\widetilde{f_{i}}$ is the forcing term. The first and
second terms on the right-hand side of Eq.~(\ref{eq:filteredKE})
represent the spatial flux of the kinetic energy and dissipation of
the energy, respectively. The term $\Pi=-\tau_{ij} \widetilde{S}_{i
  j}$ with $\tau_{ij} = \widetilde{u_i u_j} - \widetilde{u}_i
\widetilde{u}_j$ is the inter-scale kinetic energy transfer, which is
the quantity of interest here. A positive value of $\Pi$ signifies the
transfer of kinetic energy from scales above the filter cut-off to
smaller scales (forward cascade). Conversely, a negative value of
$\Pi$ indicates the transfer of kinetic energy from sub-filtered flow
motions to scales above the filter cut-off (backward cascade).  It is
worth noting that alternative definitions of $\Pi$ are possible by
rearranging the terms within the spatial flux in
Eq.~(\ref{eq:filteredKE}), and other options have been proposed in the
literature~\cite{Vela2022, cardesa2019}. Here, we adopt
$\Pi=-\tau_{ij} \widetilde{S}_{ij}$, which is one of the most widely
accepted and studied definitions within the community.
Some of the merits of $\Pi$ lie in its Galilean invariance, along with
invariance under translations and rotations. It is also easily
interpretable as the rate of transfer of kinetic energy from the
filtered motions to the residual motions, as it appears with an
opposite sign in the equation for the subfilter motions.
Additionally, $\Pi$ naturally emerges within the framework of
large-eddy simulation (LES) when considering the closure term
$\tau_{ij}$. Thus, gaining a deeper understanding of $\Pi$ could
inform the development of closure models.
\begin{figure}
    \centering
    \subfloat[]{\includegraphics[width=.45\linewidth]{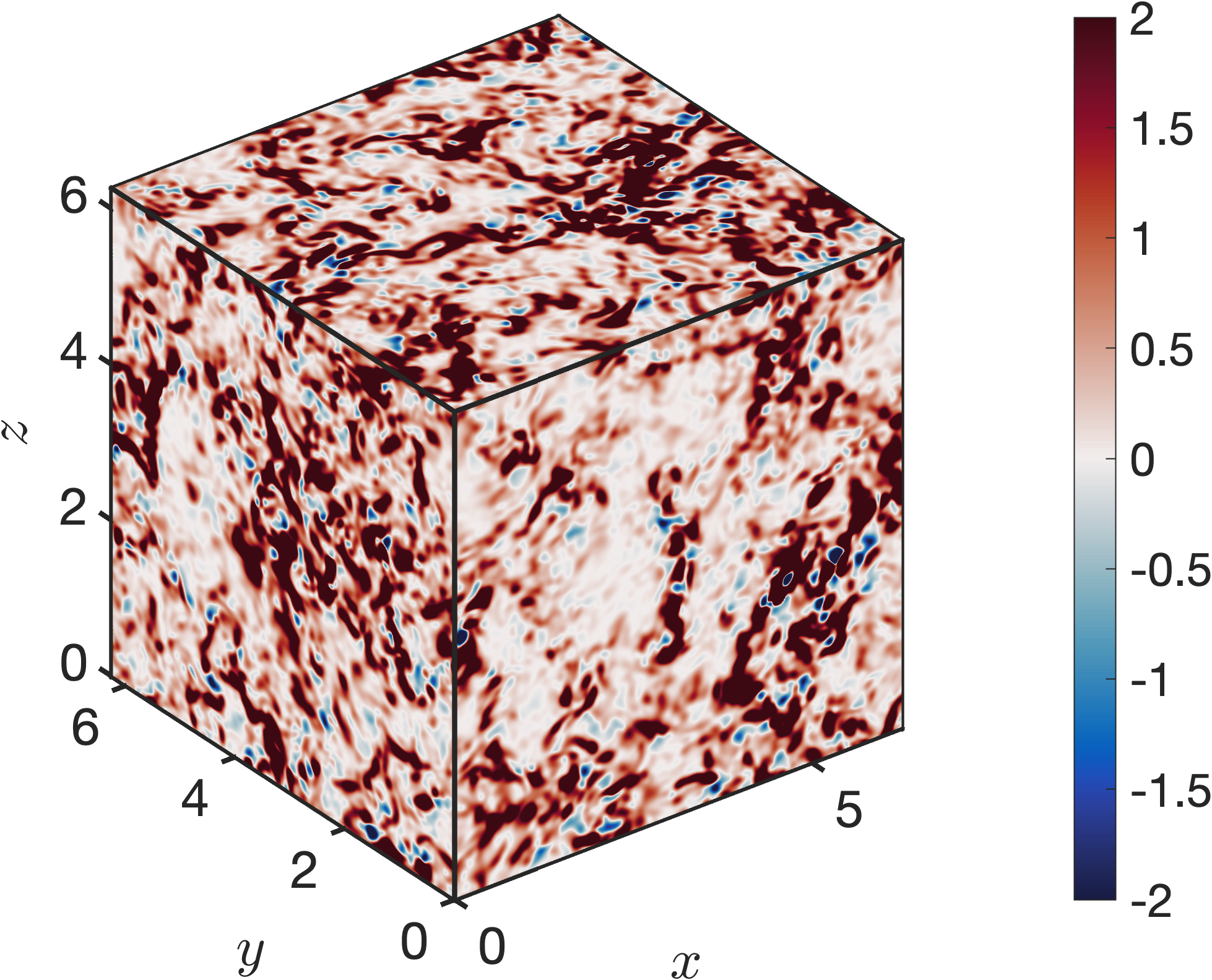}}
    \hspace{0.2cm}
    \subfloat[]{\includegraphics[width=.36\linewidth]{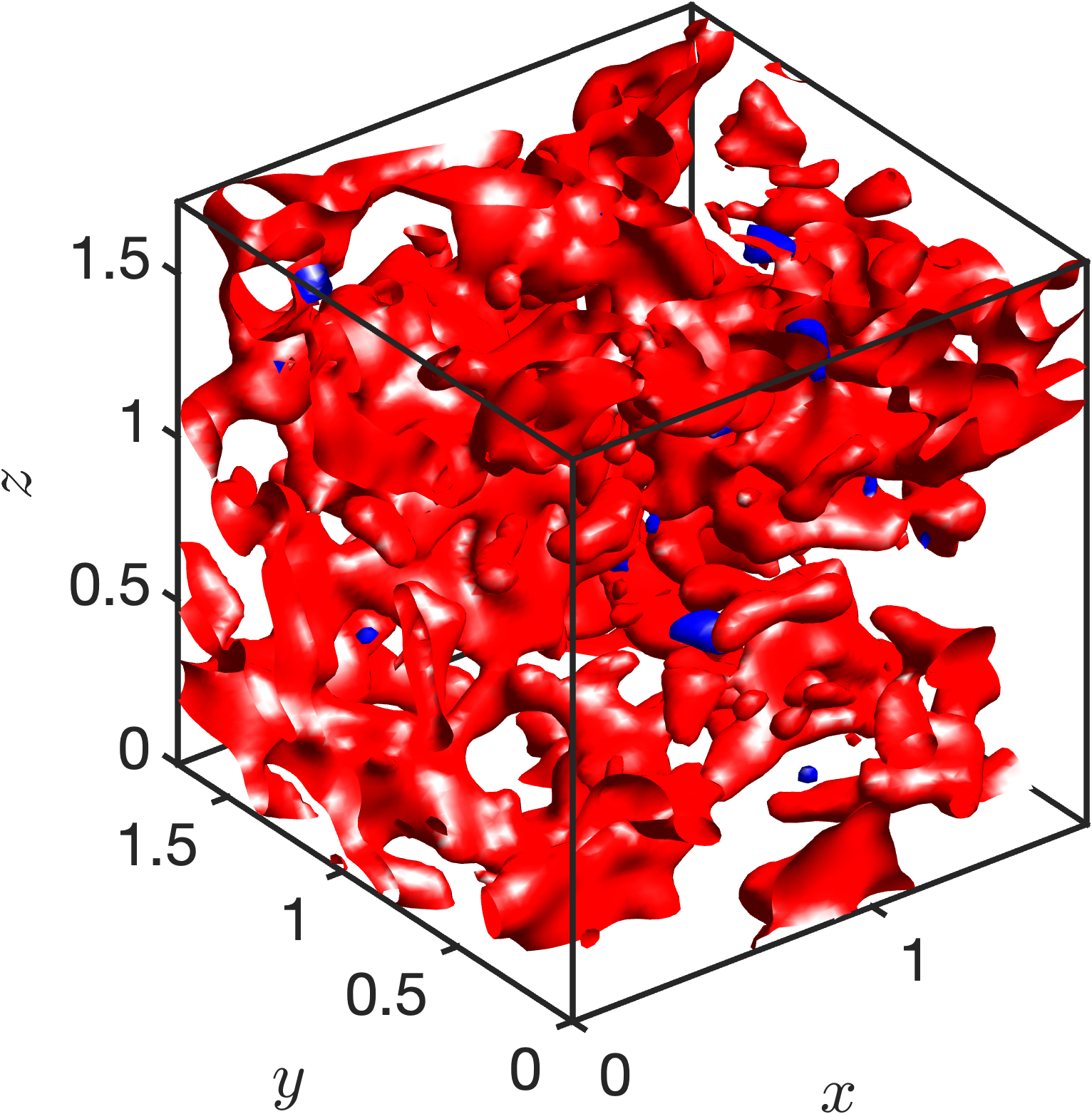}}
    \caption{Instantaneous inter-scale kinetic energy transfer. (a)
      $\Pi$ and (b) regions with $\Pi>\Pi_{\text{rms}}$ (red) and
      $\Pi<-\Pi_{\text{rms}}$ (blue). The results are for case HIT2
      and $r=30\eta$.}
    \label{fig:transfer}
\end{figure}

The instantaneous inter-scale kinetic energy transfer at one instance
is visualized in figure~\ref{fig:transfer}(a). Our primary efforts are
directed toward analyzing the coherent structure of the flow in the
vicinity of intense forward cascade events. Regions of strong positive
$\Pi$ are isolated by thresholding the field with the parameter
$\alpha > 0$, such that $\Pi > \alpha\Pi_{\text{rms}}$, where
$\Pi_{\text{rms}}$ represents the standard deviation of $\Pi$ over all
the dataset. The value of $\alpha = 1$ is selected following previous
work~\cite{Dong2020}. We show in the Supplementary Information that
the conclusions drawn in this manuscript also hold for other values of
$\alpha$.  Then, individual three-dimensional $\Pi$-structures are
defined as contiguous regions in space satisfying $\Pi >
\Pi_{\text{rms}}$. These structures will be utilized as markers to
conditionally average the flow around them.  The $\Pi$-structures
occupy less than 20\% of the total fluid volume but contribute to 70\%
of the total kinetic energy transfer.  Figure~\ref{fig:transfer}(b)
visualizes the regions associated with forward (red, $\Pi >
\Pi_{\text{rms}}$) and backward (blue, $\Pi < -\Pi_{\text{rms}}$)
intense cascade events for a given snapshot.
Note that we use $\Pi = \Pi(\mathbf{x},t)$ to designate the scalar
field of inter-scale kinetic energy transfer, which is a function of
both space and time, whereas the term $\Pi$-structure refers to the
individual structures defined by contiguous regions in space that
satisfy $\Pi(\mathbf{x},t) > \Pi_{\text{rms}}$ at a given time.

Forward cascade events dominate over their backward counterparts, as
clearly appreciated in figure~\ref{fig:transfer}(b), consistent with
previous results in the literature~\cite{ pio:cab:moin:lee:1991,
  Borue1998, Cerutti1998}.  We will focus solely on the forward energy
cascade, while we do not address inverse cascading events. This
decision is primarily driven by two factors. Firstly, the number of
intense backward cascade events is orders of magnitude smaller than
the forward events, as illustrated in
figure~\ref{fig:transfer}(b). Secondly, whereas the importance of the
forward energy cascade is widely recognized in terms of dynamics and
reduced-order modeling of the flow, the same consensus is not
established for the inverse energy cascade in three-dimensional
turbulence. Recent works have suggested that the inverse energy
cascade may play a minor role in the flow dynamics or even lack
physical significance~\cite{Vela2021, Vela2022, lozano2023,
  Martinez2024}.

\subsection*{Geometric properties of $\Pi$-structures}
\label{sec:geometry}

We characterize the geometric properties of the $\Pi$-structures in
terms of size and fractal dimensions across different filter widths
and Reynolds numbers. A local frame of reference is determined using
the center of mass and the principal axes of inertia for each
$\Pi$-structure.  \corr{The center of mass for the $n$-th
  $\Pi$-structure, denoted by $\mathbf{x}_n =
  [x_{n,1},x_{n,2},x_{n,3}]$, is computed assuming a solid object with
  constant density:
  \begin{equation}
    \mathbf{x}_n =
    \frac{\iiint_{\mathcal{V}_n} \mathbf{x}_p \mathrm{d}V_n}{\iiint_{\mathcal{V}_n} \mathrm{d}V_n},
  \end{equation}
  where $\mathbf{x}_p = [x_{p,1},x_{p,2},x_{p,3}]$ represents the
  position vector of points within the volume of the structure,
  denoted by $\mathcal{V}_n$.  The principal axes of inertia are
  obtained from the eigenvalues of the moment of inertia tensor for
  the $n$-th $\Pi$-structure:
  \begin{equation}
  I^n_{ij} = \iiint_{\mathcal{V}_n} \left( \delta_{ij} l_n^2
  - l_{n,i} l_{n,j} \right) \mathrm{d}V_n,
  \end{equation}
  where $l_{n,i} = x_{p,i}-x_{n,i}$ is the $i$-th component of the
  position vector relative to the center of mass, $l_n^2 = l_{n,i}
  l_{n,i}$ is the squared distance from the center of mass, and
  $\delta_{ij}$ is the Kronecker delta.}

\begin{figure}
  \begin{center}  
    \subfloat[]{\includegraphics[width=.46\linewidth]{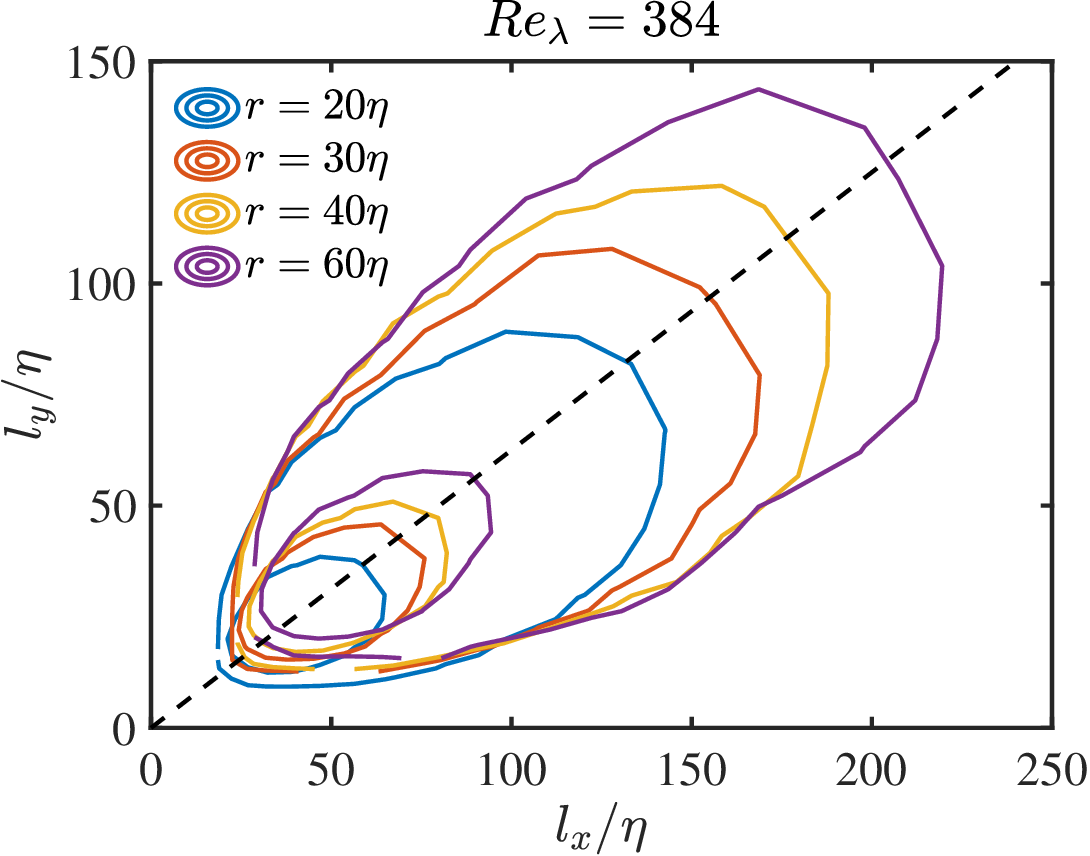}}
    \hspace{0.2cm}
    \subfloat[]{\includegraphics[width=.46\linewidth]{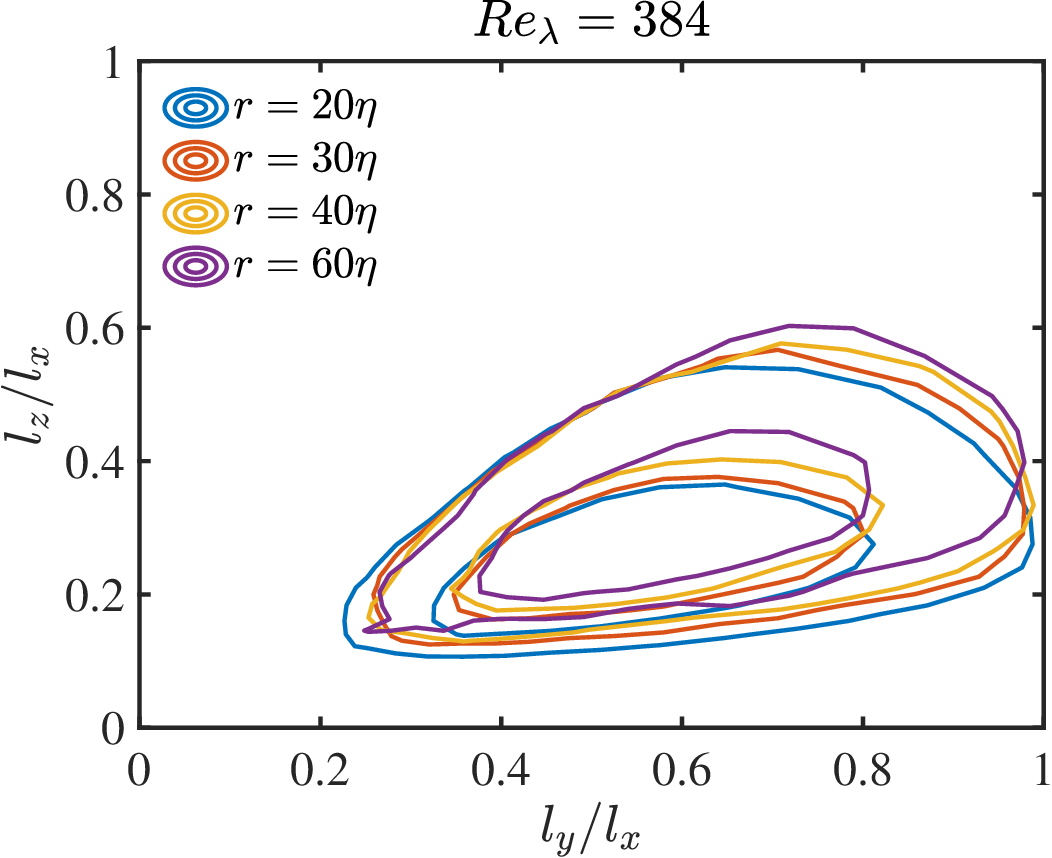}}
  \end{center}
  \begin{center}  
    \subfloat[]{\includegraphics[width=.46\linewidth]{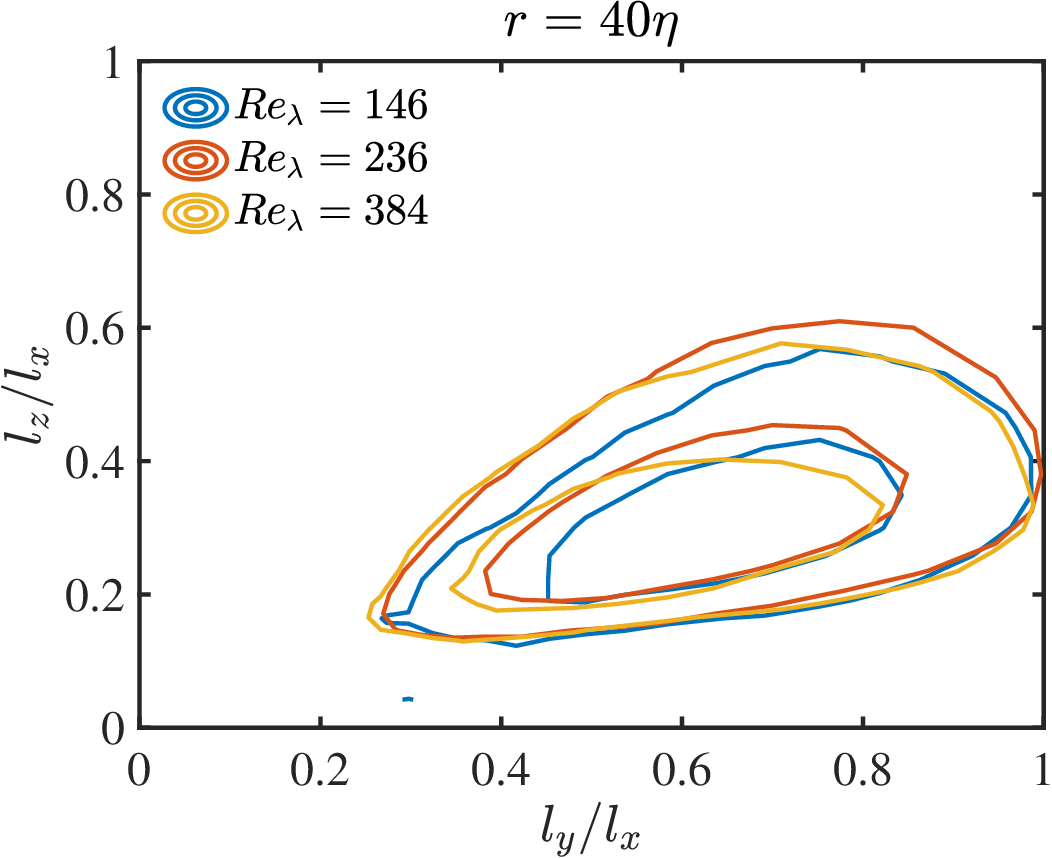}}
    \hspace{0.2cm}
    \subfloat[]{\includegraphics[width=.46\linewidth]{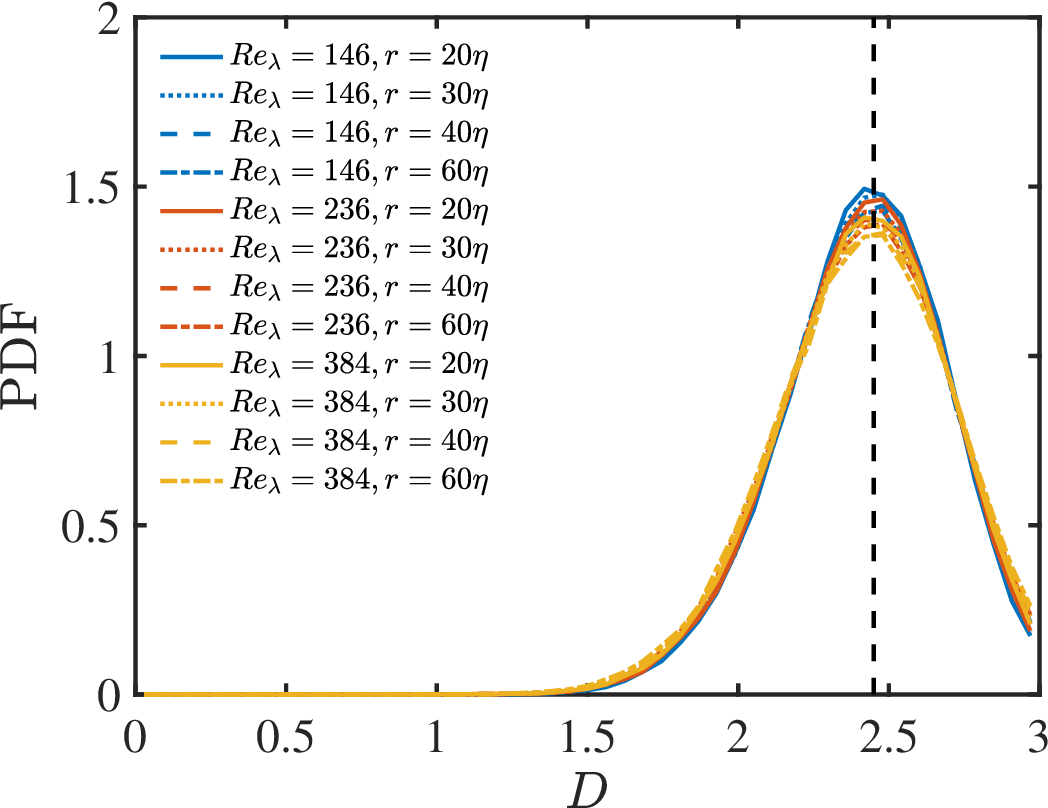}}
  \end{center}
    \caption{(a) Joint probability density function of the sizes
        $l_x$ and $l_y$ of the $\Pi$-structures for different filter
        widths. The dashed line is $l_y = l_x/1.6$. (b,c) Joint
        probability density function of the aspect ratios $l_y/l_x$
        and $l_z/l_x$ for (b) different filter widths for HIT3 and (c)
        different Re$_\lambda$ for $r=40\eta$. (d) Probability density
        function of the fractal dimension $D$ of the $\Pi$-structures
        for different filter widths and Reynolds numbers. The vertical
        dashed line is $D=2.45$.  \label{fig:size_distribution}}
\end{figure}

The three typical lengths of each $\Pi$-structure, $l_x \geq l_y \geq
l_z$, are measured by the edges of their bounding boxes aligned with
the local frame of reference.  Examples of the bounding boxes and the
circumscribed $\Pi$-structures are illustrated in
figure~\ref{fig:schematic}. The joint probability density function
(JPDF) of $l_x/\eta$ and $l_y/\eta$ is shown in
figure~\ref{fig:size_distribution}(a) for different filter widths at
Re$_\lambda=384$. The lengths follow a self-similar relationship along
$l_x \sim l_y$, i.e., longer structures tend to be proportionally
wider. Although not shown, the same self-similar relationship holds
for the third length, $l_x \sim
l_z$. Figure~\ref{fig:size_distribution}(a) shows that, as expected,
the lengths of the structures become larger with increasing values of
$r$ and that these lengths also follow the self-similar relation $l_x
\sim l_y$.  The aspect ratios of the bounding boxes are quantified in
figure~\ref{fig:size_distribution}(b,c), which features the JPDF of
$l_y/l_x$ and $l_z/l_x$ for different values of $r$ at
Re$_\lambda=384$. The results reveals that aspect ratios are similarly
distributed across scales, with values centered around $l_y/l_x
\approx 0.6$ and $l_z/l_x \approx 0.3$. Analogous aspect ratio
relationships are obtained across the three Reynolds numbers
considered (figure~\ref{fig:size_distribution}d).

The typical shape of the $\Pi$-structures is quantified via their
fractal dimension. We employed the box-counting method to calculate
the Minkowski-Bouligand dimension of each
$\Pi$-structure~\cite{Falconer2004, moi:jim:04}. For each object, the
computational domain is divided into cubes with a side length of $l$,
and the number $N_b(l)$ of boxes containing at least one point of the
object is counted. In the case of a pure fractal set with dimension
$D$, the number of boxes would follow a power law $N_b(l) \sim
l^{-D}$. In practice, this relationship only holds within a very
restricted range of scales, bounded by a large-scale and a small-scale
cut-off. We define the local fractal dimension as
  \begin{equation}
    D_l(l) = - \frac{\mathrm{d} \ln N_b(l)}{\mathrm{d} \ln l}.
  \end{equation}
The fractal dimension of the object can be taken along the range over
which the slope of $D_l(l)$ is minimum or approximately constant.

Figure~\ref{fig:size_distribution}(d) shows the PDF of the fractal
dimensions for the $\Pi$-structures for different filter widths and
Re$_\lambda$. Space-filling spheroidal shapes would exhibit fractal
dimensions close to $D \approx 3$, whereas sheet-like or filament-like
structures would have $D \approx 2$ or $D \approx 1$,
respectively. The PDF of $D$ reveals a bell-shaped distribution
centered around $2.45$, suggesting that the objects have an
intermediate form between sheet-like and strictly spheroidal
shapes. Therefore, although some $\Pi$-structures may exhibit complex
shapes, most often they resemble elongated ellipsoids with arbitrary
orientation. In summary, the results from this section show that
$\Pi$-structures are geometrically self-similar across scales within
the inertial range and for the three Re$_\lambda$ considered.  This
finding justifies the conditional averaging procedure introduced in
the subsequent section.
  
\subsection*{Conditional-averaged flow fields}
\label{sec:conditional}

\begin{figure}
  \centerline{\includegraphics[width=1\linewidth]{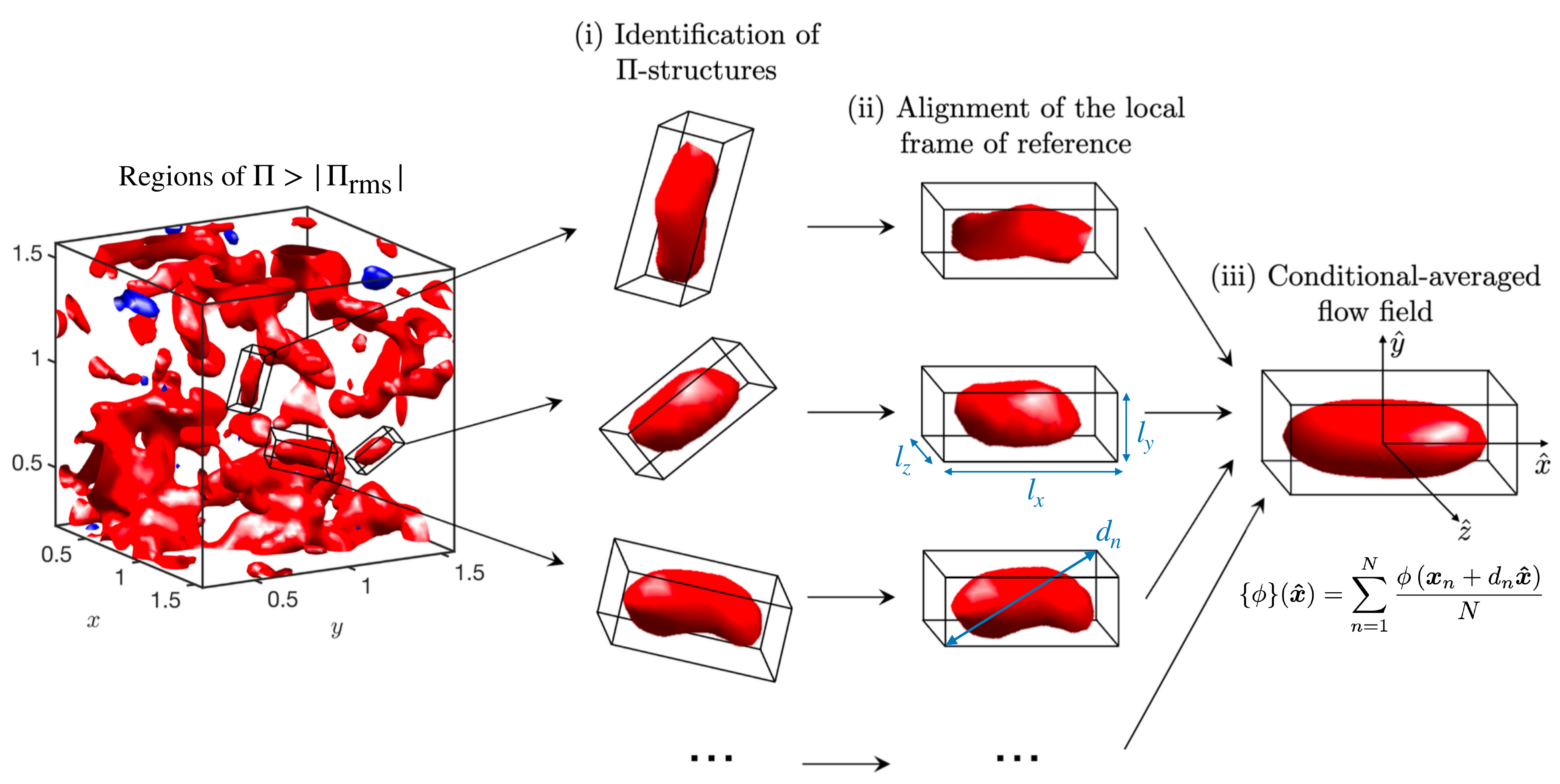}}
  \caption{Schematic of the conditional averaging procedure. The
    normalized directions of the local frame of reference attached to
    the principal axes of inertia of each $\Pi$-structure are
    $[\hat{x}, \hat{y}, \hat{z}]$.  The three lengths of the bounded
    box are $l_x$, $l_y$ and $l_z$. The diagonal of the bounding box
    for the $n$-th $\Pi$-structures is denoted by $d_n$. The
    conditional-averaged quantity $\phi$ is represented by
    $\{\phi\}$.}
\label{fig:schematic}
\end{figure}

We calculate the average flow surrounding $\Pi$-structures accounting
for symmetries and self-similarity. Figure~\ref{fig:schematic}
presents a schematic of the conditional averaging procedure. This
process involves three steps, illustrated in the figure from left to
right: identification of the $\Pi$-structures, calculation of the
local frame of reference, and computation of the conditional-averaged
flow field. Each step is detailed below.
\begin{enumerate}
\item[(i)] First, individual $\Pi$-structures are identified using the
  thresholding approach previously discussed. Examples of three
  individual $\Pi$-structures are shown in
  figure~\ref{fig:schematic}. A characteristic length scale is
  assigned to each $\Pi$-structure, defined by the diagonal of their
  bounding box, $\sqrt{l_x^2 + l_y^2 + l_z^2 }$. Our interest lies in
  studying structures that are of a size similar to the filter
  width. $\Pi$-structures larger than half the length of the entire
  computational domain are excluded from the analysis. Similarly,
  structures smaller than $3^3$ grid points are also disregarded. The
  number of valid $\Pi$-structures considered for the analysis is
  roughly on the order of $\mathcal{O}(10^4)$ for each Re$_\lambda$
  and filter width.  Given the large number of samples, the
  statistical uncertainty of the results were found to be small and
  more details are discussed in the Supplementary Information.
\item[(ii)] Each individual $\Pi$-structure is assigned a local frame
  of reference, with the origin positioned at its center of mass.
  This is the same frame of reference used in the previous section to
  compute the characteristic lengths of the $\Pi$-structures.  The
  axes directions are aligned with the principal axes of inertia of
  the structure relative to the center of mass.  A rotation matrix,
  constructed from these eigenvectors, is used to rotate the flow
  field surrounding each structure. Figure~\ref{fig:schematic}
  illustrates the rotated bounding boxes for three $\Pi$-structures.
  The principal axes of inertia introduced above does not inherently
  define the positive or negative directions of the axes. To resolve
  this ambiguity, we set the direction of the axes such that the
  highest enstrophy $\widetilde{\omega}_i \widetilde{\omega}_i$,
  averaged over each of the eight quadrants of the bounding box, is
  located in the first quadrant.
\item[(iii)] Conditional-averaged quantities are computed by ensemble
  averaging over the $\Pi$-structures in the local frame of
  reference. The conditional-averaged is performed over the
  $\Pi$-structures for a given Re$_\lambda$ and $r$. Prior to
  performing the average, the spatial coordinates of the local frame
  of reference attached to each individual structure are re-scaled
  based on the diagonal of the rotated bounding box of the
  $\Pi$-structure. This process is grounded in the geometric
  self-similarity of the $\Pi$-structures discussed in above. The
  conditional-averaged $\phi$, denoted by $\{\phi\}$, is calculated as
\begin{equation}
\{\phi\}(\mathbf{\hat{x}})=\sum_{n=1}^{N}
\frac{\phi\left(\mathbf{x}_{n}+d_{n}
  \mathbf{\hat{x}}\right)}{N},
\end{equation}
where $n=1, ..., N$ is the label for each $\Pi$-structure (with $N$
the total number of samples), $\mathbf{x}_n$ is the center of mass
of the $n$-th $\Pi$-structure, $d_n$ is the diagonal length of
bounding box of the $n$-th $\Pi$-structure, and $\mathbf{\hat{x}}$
is the space coordinate relative to the local frame of reference,
re-scaled by the diagonal length.

The conditional average approach outlined here enables the study of
local-in-space flow patterns surrounding cascading events. We have
avoided traditional spectral analysis due to its global-in-space
nature, which renders the approach unsuitable for elucidating the
structure of localized events. Note that the conditional average
procedure is not merely a qualitative description of the flow, but
rather a quantitative representation of the most probable states and
patterns in the flow surrounding cascading events.

\end{enumerate}

\section*{Results}
\label{sec:results}

\subsection*{Conditional-averaged flow surrounding intense cascade events}

\begin{figure}
    \centering
    \subfloat[]{\includegraphics[width=.48\linewidth]{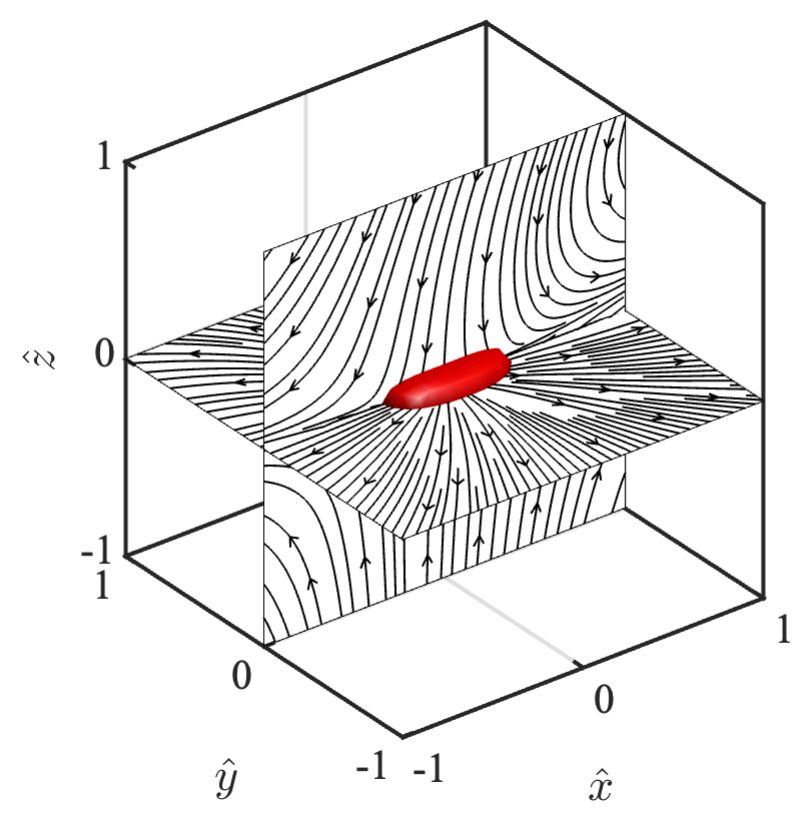}}
    \subfloat[]{\includegraphics[width=.48\linewidth]{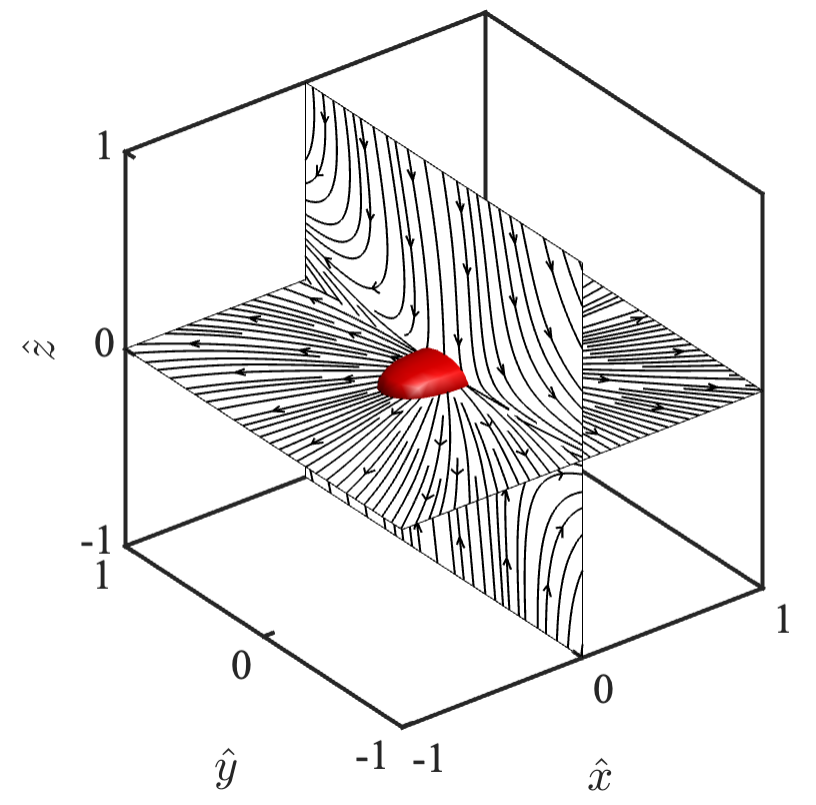}}
    \caption{Streamlines of the conditional-averaged velocity
      ($\{\tilde{u}\}, \{\tilde{v}\}, \{\tilde{w}\}$) surrounding
      intense energy cascade events. The streamlines are constrained
      to the planes (a) $\hat{y}=0$ and $\hat{z}=0$ (b) $\hat{x}=0$
      and $\hat{z}=0$. The red isosurface is 50\% probability of
      finding a point belonging to a $\Pi$-structure. The results are
      for case HIT3 and $r=30\eta$.\label{fig:velocity_3d}}
\end{figure}

The dominant velocity patterns surrounding intense energy transfer
events are shown in figure~\ref{fig:velocity_3d}. The region of
intense energy transfer is represented by a red iso-surface,
corresponding to 0.5 of the maximum probability of finding a point
belonging to the $\Pi$-structure, i.e., $P(\Pi > \Pi_{\text{rms}}) =
0.5$. The averaged shape is an elongated ellipsoid, consistent with
our geometric analysis.  The streamlines are calculated from the
conditional-averaged velocity field ($\{\tilde{u}\}, \{\tilde{v}\},
\{\tilde{w}\}$) in the vicinity of the $\Pi$-structures after
subtracting the mean velocity at the origin of the $\Pi$-structure.
Notably, the flow exhibits a saddle point topology centered at the
origin of the $\Pi$-structure, which is a characteristic signature of
strain-dominated regions.  Similar patterns have previously been
reported in the context of sweep and ejection pairs in homogeneous
shear turbulence~\cite{Dong2020} and wall-bounded
turbulence~\cite{Natrajan2006, Hong2012}. However, the results
presented here specifically pertain to the inertial range of isotropic
turbulence.

\begin{figure}
    \begin{center}
      \subfloat[]{\includegraphics[width=.48\linewidth]{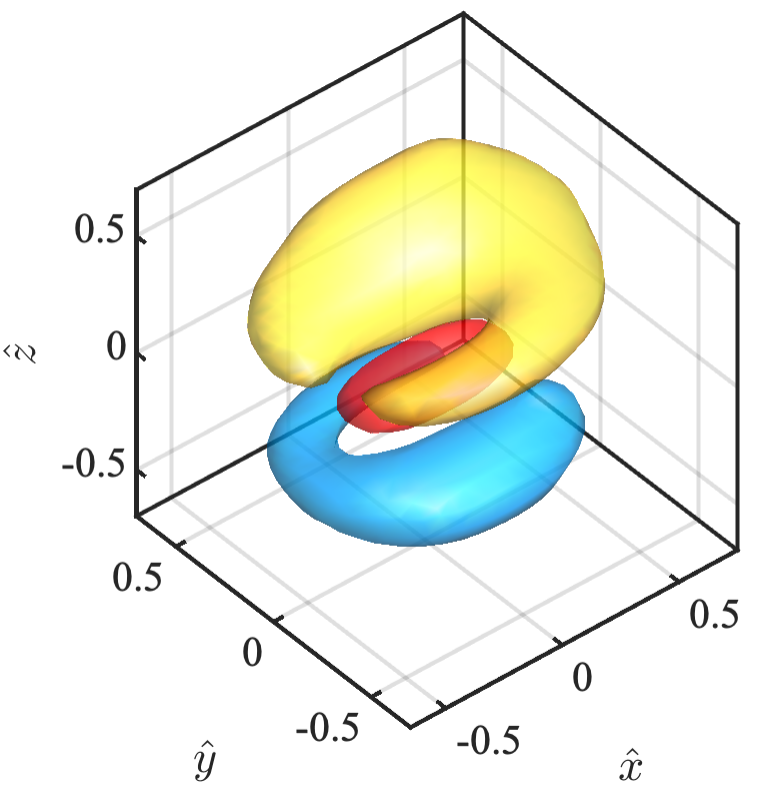}}
      \subfloat[]{\includegraphics[width=.48\linewidth]{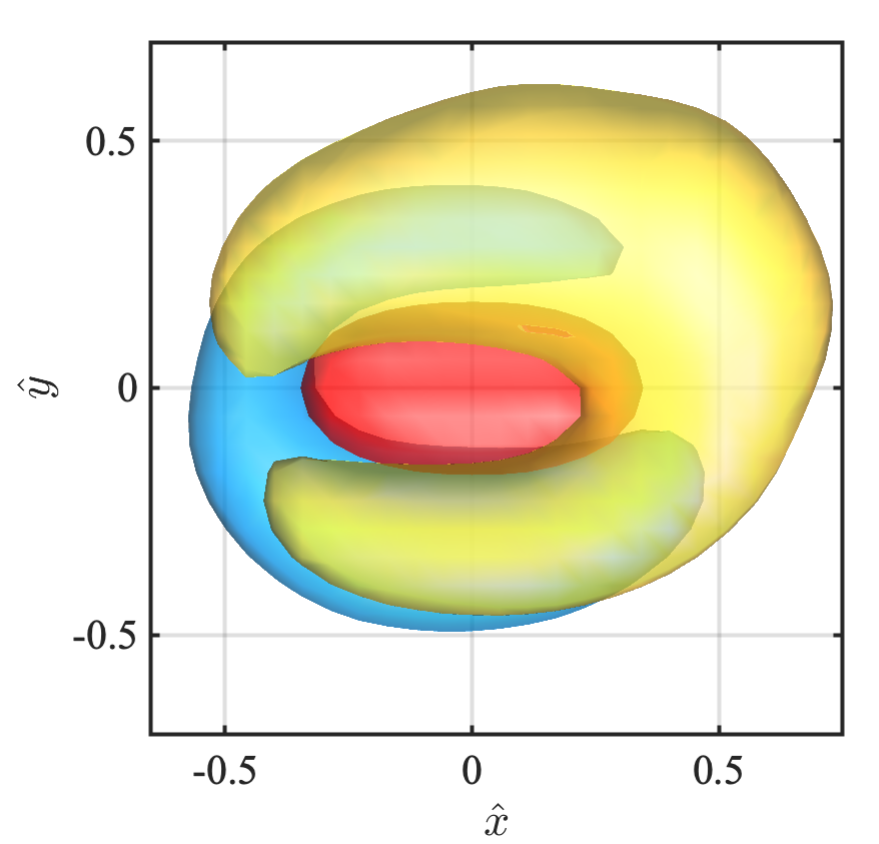}}
    \end{center}
        \begin{center}
      \subfloat[]{\includegraphics[width=.51\linewidth]{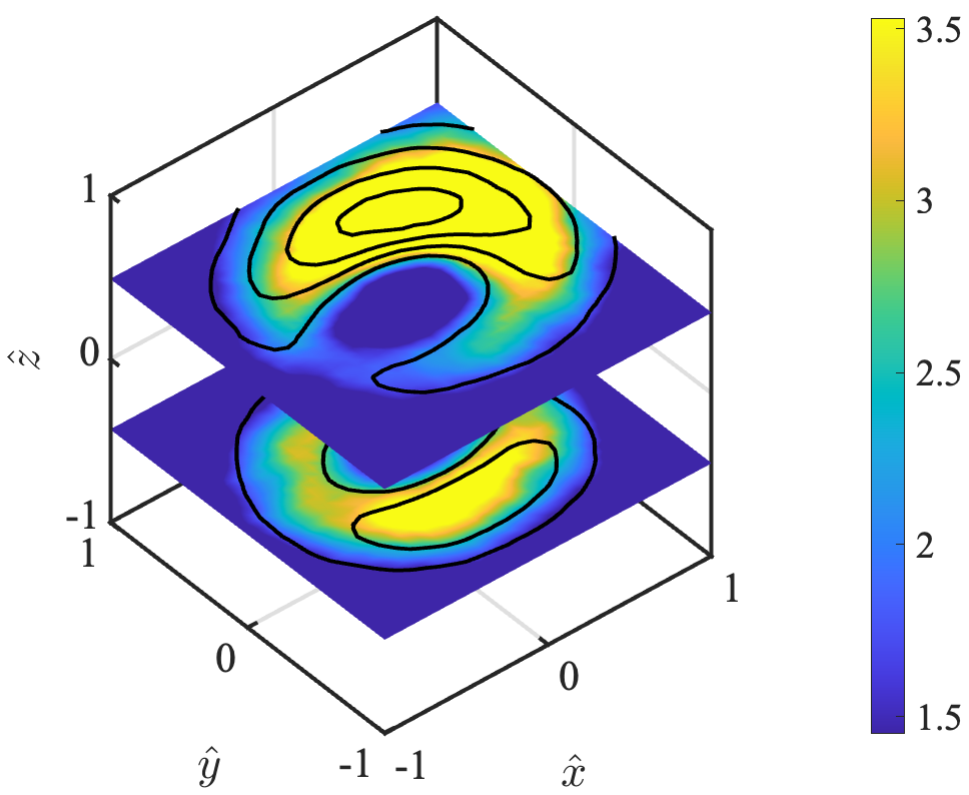}}
      \subfloat[]{\includegraphics[width=.49\linewidth]{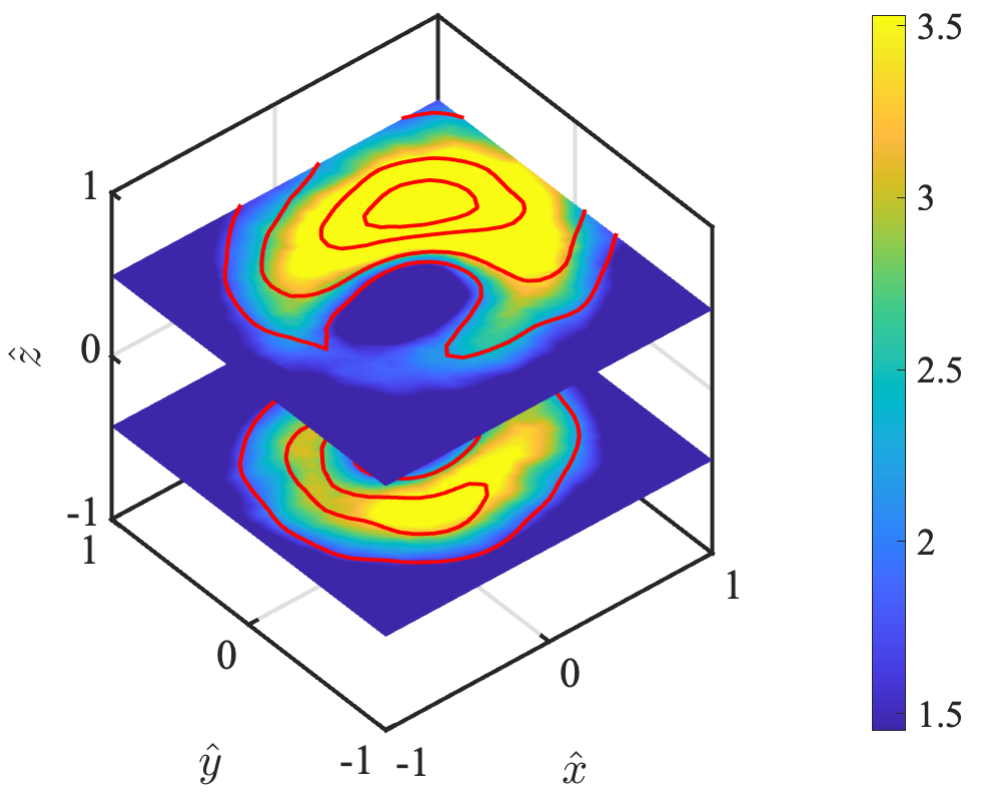}}
    \end{center}
    \caption{(a,b) Iso-surface of the probability of finding a point
      belonging to a $\Pi$-structure equal to 0.5 (red) and
      conditional-averaged enstrophy
      $\{\widetilde\omega_i\}\{\widetilde\omega_i\}$ at $0.35\times
      \max\left(\{\widetilde\omega_i\}^2\right)$ at $\hat{z}>0$
      (yellow) and $\hat{z}<0$ (blue). The results are shown in
      isometric view in panel (a) and in the $\hat{x}$-$\hat{y}$ view
      in panel (b).  The results are for case HIT3 and
      $r=30\eta$. (c,d) Conditional-averaged enstrophy for (c) HIT3
      and $r=30\eta$ (color) and HIT2 and $r=30\eta$ (solid black
      line); (d) HIT3 and $r=30\eta$ (color) and HIT3 and $r=40\eta$
      (solid red line). The values are normalized by the standard
      deviation of
      $\{\widetilde\omega_i\}\{\widetilde\omega_i\}$.  \label{fig:enstrophy}}
\end{figure}

The enstrophy pattern ($\{\widetilde\omega_i\}\{\widetilde\omega_i\}$)
surrounding intense energy transfer events is visualized in
figure~\ref{fig:enstrophy}. The red region in
figure~\ref{fig:enstrophy}(a,b) represents again the 0.5 probability
iso-surface for locating a point belonging to a $\Pi$-structure. The
iso-surfaces of the enstrophy field are set at 35\% of the maximum
value. To facilitate the visualization of enstrophy, the regions for
$\hat{z} > 0$ and $\hat{z} < 0$ are colored in yellow and blue,
respectively. The distinctive emerging pattern reveals that the energy
cascade occurs between two regions of intense enstrophy.  These
regions manifest as hairpin-like shapes with opposing orientations
that do not overlap with the region where the energy transfer reaches
its maximum. The enhanced enstrophy located at $\hat{x} > 0$, $\hat{y}
> 0$, and $\hat{z} > 0$, is due the conditional average procedure,
which rotates the local frame of reference to favor higher enstrophy
within the first quadrant. However, the existence of two enstrophy
regions, the specific shapes of these regions, and the absence of
overlap with the $\Pi$-structures are not the result of any
constraint, but rather a manifestation of the statistical significance
of that arrangement.

The robustness of the results across Re$_\lambda$ and filter widths is
assessed in figures~\ref{fig:enstrophy}(c) and (d). The figures
display the colormap of the conditional-averaged enstrophy in two
planes cutting along the hairpin-like structures identified in
figures~\ref{fig:enstrophy}(a,b). This approach complements that of
figure~\ref{fig:enstrophy} by eliminating the need to select a
threshold to visualize the hairpin-like shape. The solid contours
represent the conditional-averaged enstrophy for a different
Re$_\lambda$ (figure~\ref{fig:enstrophy}c) or different filter width
(figure~\ref{fig:enstrophy}d). Some small sensitivities can be
observed with respect to Re$_\lambda$ and filter width, likely due to
low Reynolds number effects. Nonetheless, the findings are consistent
with those from figures~\ref{fig:enstrophy}(a,b) and the same
conclusions are drawn.

\begin{figure}
    \centering
    \subfloat[]{\includegraphics[width=.48\linewidth]{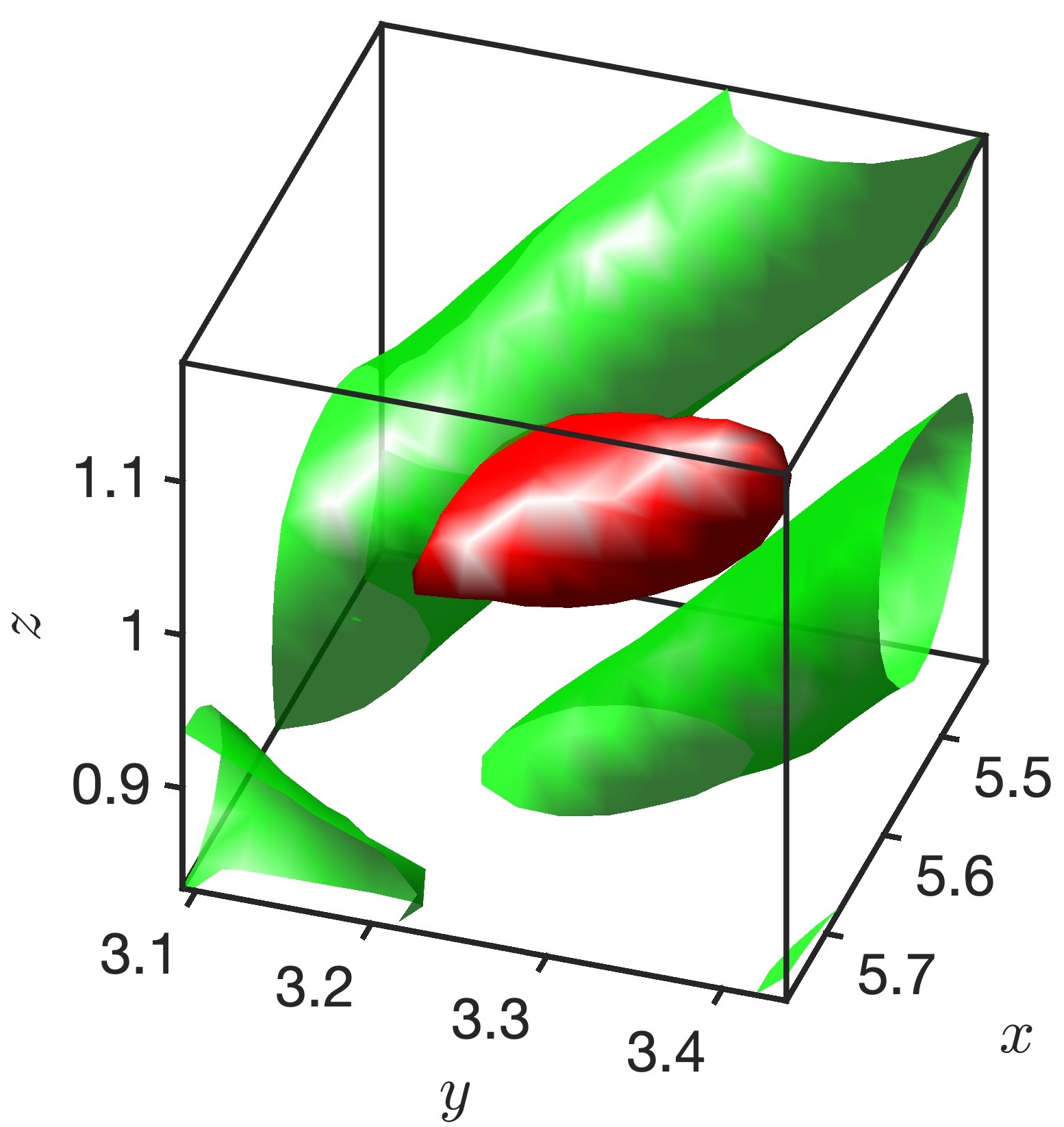}}
    \hspace{0.1cm}
    \subfloat[]{\includegraphics[width=.48\linewidth]{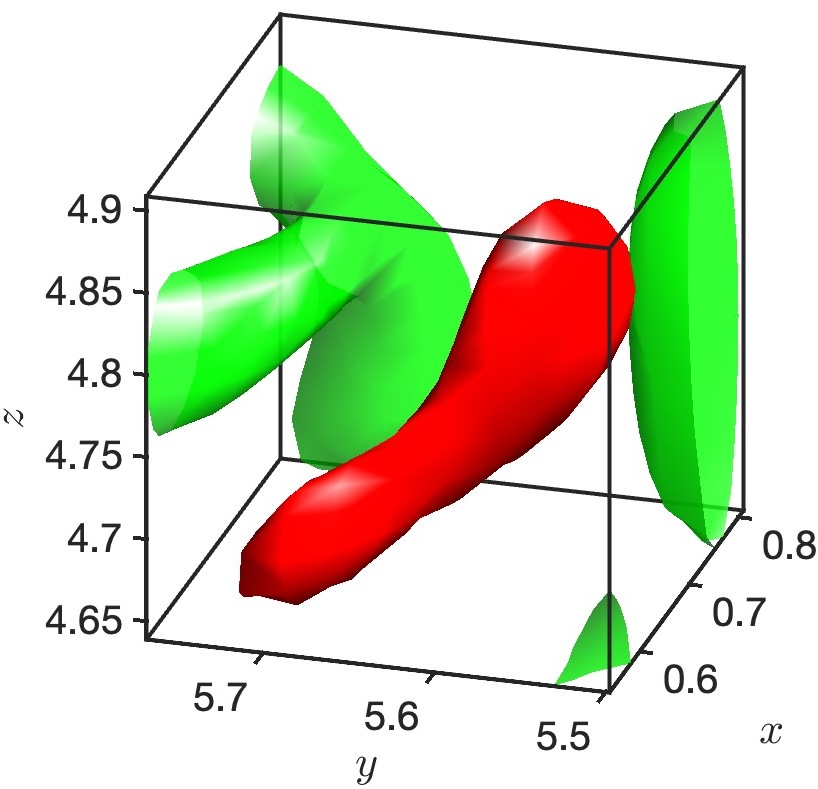}}
  \caption{Examples of instantaneous intense cascade events:
    iso-surfaces of high transfer regions $\Pi>\Pi_{\text{rms}}$ (red)
    and high enstrophy regions at $0.025 \times \max
    \left(\widetilde\omega_i \widetilde\omega_i\right)$ (green). The
    results are for case HIT2 and $r=30\eta$.}
    \label{fig:instantaneous}
\end{figure}

It is important to note that the hairpin-like structure identified
does not imply that instantaneous vortices are shaped in that
manner. Rather, it signifies that the irregular arrays of vortices
tend to be preferentially located in those regions.  Two examples of
the instantaneous field surrounding the $\Pi$-structures are depicted
in figure~\ref{fig:instantaneous}. It can be seen that intense energy
transfer events ($\Pi > \Pi_{\text{\text{rms}}}$, colored in red) are
surrounded by areas of high enstrophy ($\widetilde{\omega}_i
\widetilde{\omega}_i$, colored in green). In these particular
examples, multiple vortex tubes are positioned closely to the
$\Pi$-structures, yet without overlapping in space. Similar instances,
consisting of regions of intense $\Pi$ surrounded by distinct regions
of intense $\widetilde{\omega}_i \widetilde{\omega}_i$, were
consistently observed. While certain $\Pi$-structures did overlap with
vortex tubes, such occurrences were less frequent, in accordance with
the results from the conditional-averaged flow.

\begin{figure}
  \centering
  \subfloat[]{\includegraphics[width=.33\linewidth]{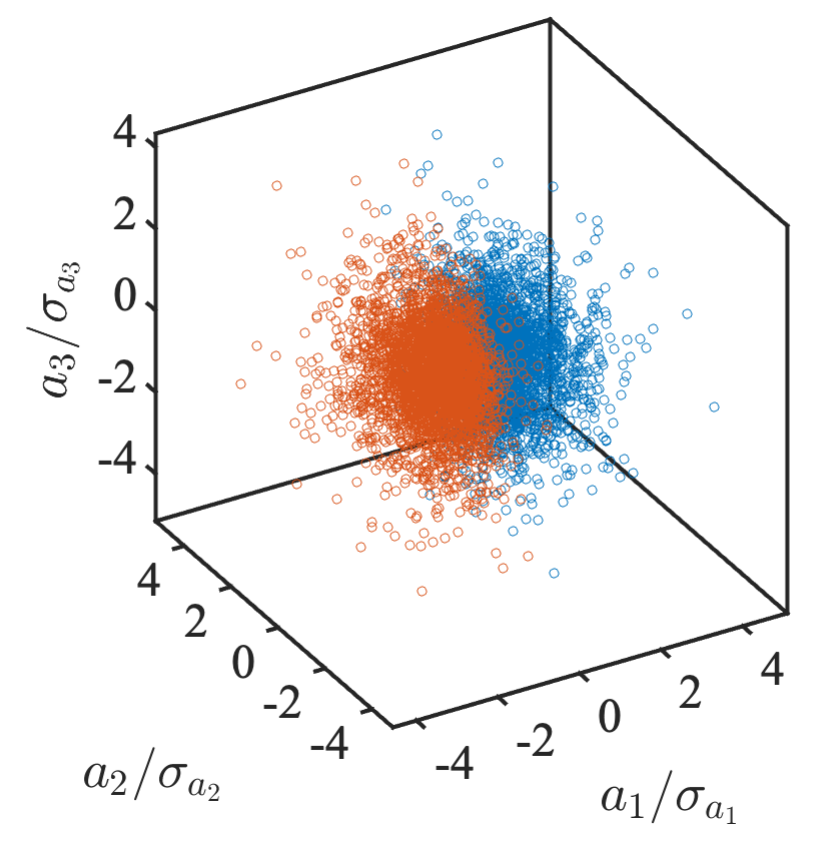}}
  \subfloat[]{\includegraphics[width=.33\linewidth]{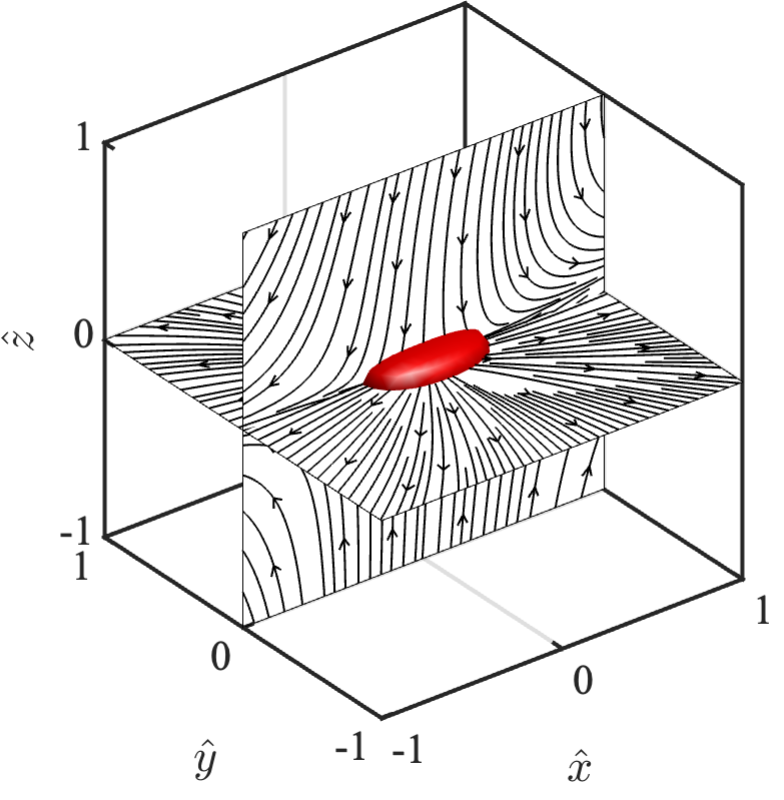}}
  \subfloat[]{\includegraphics[width=.33\linewidth]{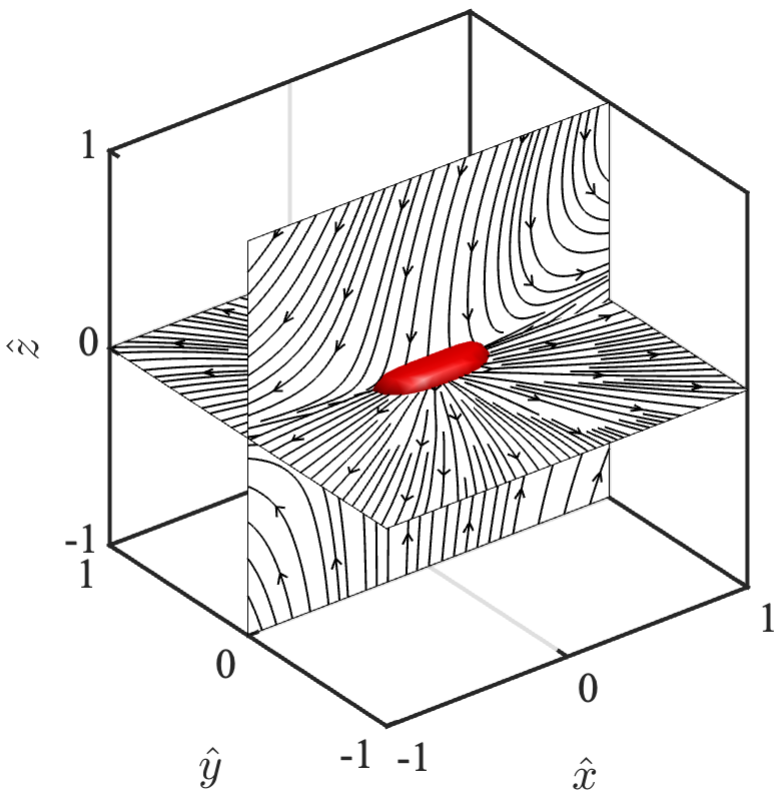}}
  \caption{(a) Point cloud of the three most energetic POD
    coefficients associated with the velocity field around
    $\Pi$-structures. The coefficients are normalized by their
    standard deviation and colored according to the k-mean group they
    belong to. (b,c) Conditional-averaged shape of transfer object and
    nearby averaged velocity streamlines for (b) group 1 and (c) group
    2. The contours represent 0.5 of the maximum probability of
    finding a $\Pi$-structure. The streamlines are constrained to the
    planes $\hat{y}=0$ and $\hat{z}=0$. \label{fig:kmeans}}
\end{figure}

The results above rely on the assumption that only one type of flow
pattern surrounds the intense energy transfer events. However, if
patterns with different flow topologies are involved in the cascading
process, performing an ensemble average over all the $\Pi$-structures
would yield a distorted image of the coherent structure. To assess the
possibility of distinct flow patterns contributing to the energy
cascade, we divide the samples used to compute $\{\phi\}$ into $n$
groups with the goal of separating instances containing different flow
structures.
\corr{ The goal is to create groups of samples that are as distinct as
  possible. This is accomplished by first extracting the large-scale
  velocity patterns surrounding the $\Pi$-structures using proper
  orthogonal decomposition (POD), followed by applying the k-means
  algorithm to maximize the differences between samples in different
  groups. The POD modes are computed for the velocity vector
  $[\tilde{u}, \tilde{v}, \tilde{w}]$ after subtracting the mean
  velocity at the origin of the $\Pi$-structure. The spatial extent of
  the domain for computing the POD modes is defined by a cube with a
  side length of 1 and origin at the center of mass of the
  structure. As shown in figure~\ref{fig:enstrophy}, this domain is
  sufficient to enclose the key flow patterns. The first three most
  energetic POD modes are used to characterize the large-scale flow
  pattern for each $\Pi$-structure. The corresponding POD coefficients
  are denoted as $a_1(n)$, $a_2(n)$, and $a_3(n)$, where $n$ refers to
  the sample (i.e., the $\Pi$-structure) they belong to.}
The k-means algorithm is applied to place the samples into $n$ groups
that minimize within-cluster variances of the three most energetic POD
coefficients. The POD coefficients and groups are shown in
figure~\ref{fig:kmeans}(a) for the case $n=2$. The
conditional-averaged velocity field for each group is portrayed in
figures~\ref{fig:kmeans}(b) and (c). Comparison of the results
indicates that both groups share similar flow features despite the
division of the samples being targeted to separate velocity patterns.
Although not shown, the process was repeated for $n=3$ and $n=4$
groups, which yielded similar conclusions. These findings suggest that
the conditional-averaged flow structure outlined in this section is
physically relevant and not merely an artifact of the averaging
process. Additionally, the results in figure~\ref{fig:kmeans} are
computed for case HIT2 and $r=30\eta$, which demonstrates that the
streamlines in figure~\ref{fig:velocity_3d}(a) also hold for different
Re$_\lambda$ and filter widths.

\subsection*{Mechanisms responsible for the inter-scale energy transfer}
\label{sec:mech}


To gain further insight into the underlying mechanisms driving the
energy cascade, we investigate the contributions of the vortex
stretching and strain-rate self-amplification in the vicinity of
intense energy transfer events. To that end, the kinetic energy
transfer $\Pi$ is decomposed into terms ascribed with different
mechanisms.  The approach adopted here follows the work by
Ref.~\cite{Johnson2020}.  We outline the key aspects of the
decomposition and the reader is referred to the original work for more
details. First, let us denote the low-pass Gaussian filter at scale
$r$ as $\overline{(\cdot)}^r$ such that $\tau_{ij}^r = \overline{u_i
  u_j}^r - \overline{u_i}^r\overline{u_j}^r$. In previous sections, we
have referred to $\overline{(\cdot)}^r$ as $\widetilde{(\cdot)}$.  The
decomposition of $\Pi$ is grounded on the mathematical relationship
between the Gaussian-filter $\tau_{ij}^r$ and the diffusion equation
  \begin{equation}
    \label{eq:GK}
    12\frac{\partial \tau_{ij}^r }{\partial (r^2) } = \frac{1}{2}\nabla^2 \tau_{ij}^r + \overline{A}_{ik}^r \overline{A}_{jk}^r, \quad \tau_{ij}^{r=0} = 0,   
  \end{equation}
where $A_{ij}=\partial u_i/\partial x_j$ is the velocity gradient
tensor and $\nabla^2$ is the Laplacian operator. By solving
Eq.~(\ref{eq:GK}) and projecting onto $\widetilde{S}_{ij}$, the
kinetic energy transfer can be decomposed as
\begin{equation}\label{eq:GK_terms}
    \Pi = \Pi_{l}^S + \Pi_{l}^\omega+ \Pi_{nl}^S + \Pi_{nl}^\omega + \Pi_{nl}^C.
\end{equation}
The terms above are given by
\begin{eqnarray}
  \Pi^S_l &=& -\frac{r^2}{12} \overline{S}_{ij}^r \overline{S}_{jk}^r \overline{S}_{ki}^r, \\
  \Pi^\omega_l &=& \frac{r^2}{48} \overline{\omega}_i^r \overline{\omega}_j^r \overline{S}_{ij}^r, \label{eq:Pi_1}\\
  \Pi^S_{nl} &=& - \frac{1}{12} \int_0^{r^2} \left(
  \overline{\overline{S}_{ik}^\theta  \overline{S}_{jk}^\theta}^{\phi}-
    \overline{\overline{S}_{ik}^\theta}^{\phi} \overline{\overline{S}_{jk}^\theta}^{\phi}
  \right) \overline{S}_{ij}^r\mathrm{d}\theta^2, \label{eq:Pi_2}\\
  \Pi^\omega_{nl} &=&  \frac{1}{48} \int_0^{r^2} \left(
  \overline{\overline{\omega}_i^\theta  \overline{\omega}_j^\theta}^{\phi}-
    \overline{\overline{\omega}_i^\theta}^{\phi} \overline{\overline{\omega}_j^\theta}^{\phi}
    \right) \overline{S}_{ij}^r\mathrm{d}\theta^2,\label{eq:Pi_3}\\
    \Pi^C_{nl} &=& - \frac{1}{12} \int_0^{r^2} \left(
  \overline{\overline{S}_{ik}^\theta  \overline{\Omega}_{jk}^\theta}^{\phi} + 
    \overline{\overline{S}_{ik}^\theta}^{\phi} \overline{\overline{\Omega}_{jk}^\theta}^{\phi}
  \right) \overline{S}_{ij}^r\mathrm{d}\theta^2, \label{eq:Pi_4}
\end{eqnarray}
where $\phi = \sqrt{r^2-\theta^2}$ and $\Omega_{ij} = (\partial
u_i/\partial x_j - \partial u_j/\partial x_i )/2$ is the
rate-of-rotation tensor.

The terms $\Pi_{l}^S$ and $\Pi_{l}^\omega$ (where $l$ denotes `local')
are the energy transfer due to local-in-scale interactions of
$\overline{S}_{ij}^r$ with either the rate-of-strain tensor
$\overline{S}_{ij}^r$ (i.e., strain self-amplification) or the
vorticity vector $\overline{\omega}_{i}^r$ (i.e, vortex stretching),
respectively.  The terms $\Pi^S_{nl}$ and $\Pi^\omega_{nl}$ (where
$nl$ denotes `non-local') are analogous to $\Pi_{l}^S$ and
$\Pi_{l}^\omega$ but correspond to non-local-in-scale interactions.
Despite the non-local nature of $\Pi^S_{nl}$ and $\Pi^\omega_{nl}$, it
has been shown that they still represent interactions occurring close
in scale~\cite{Domaradzki1990, Cardesa2015, Johnson2020,
  Johnson2021, lozano2023}. The remaining term, $\Pi_{nl}^C$ is also
non-local in scale and represents energy transfer by the resolved
strain-rate tensor acting on the product of strain-rate and vorticity.

\begin{figure}
  \centering
  \subfloat[]{\includegraphics[width=.48\linewidth]{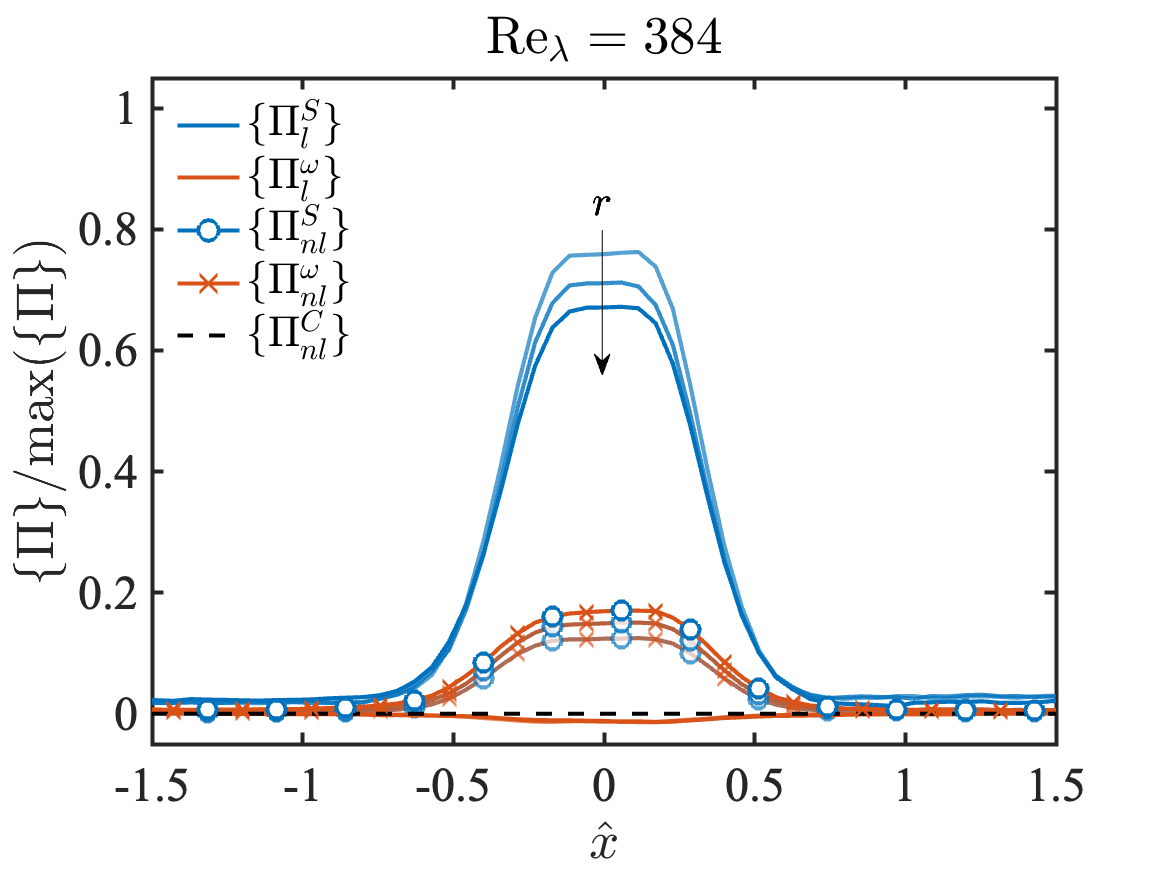}}
  \subfloat[]{\includegraphics[width=.48\linewidth]{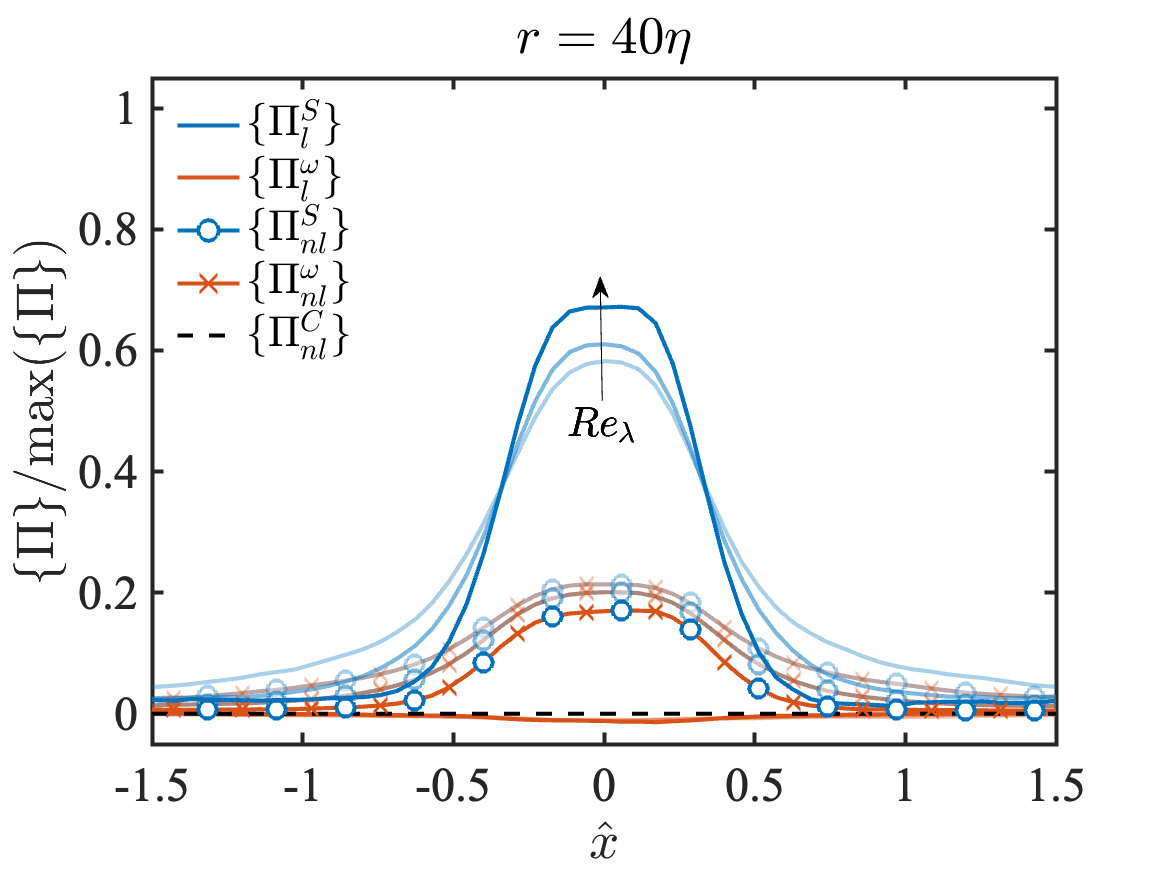}}
    \caption{ Conditional-averaged contributions to $\{\Pi\}$
      surrounding intense energy-transfer events. The results are
      shown along the $\hat{x}$ coordinate for
      $\hat{y}=\hat{z}=0$. The terms are normalized with the maximum
      of $\{\Pi\}$. (a) Contributions for HIT3 and $r/\eta=20,30$, and
      $40$ (from light to dark). (b) Contributions for $r/\eta=40$ and
      HIT1, HIT2, and HIT3 (from light to dark).}
    \label{fig:Pi}
\end{figure}

The conditional-averaged values for the total energy transfer
($\{\Pi\}$) and its individual contributions ($\{\Pi_{l}^S\}$,
$\{\Pi_{l}^\omega\}$, $\{\Pi_{nl}^S\}$, $\{\Pi_{nl}^\omega\}$, and
$\{\Pi_{nl}^C\}$) are calculated following the ensemble average
procedure from figure~\ref{fig:schematic}. For ease of visualization,
the values are plotted in figure~\ref{fig:Pi} along the $\hat{x}$
coordinate, but a similar picture emerges along the other directions,
$\hat{y}$ and $\hat{z}$. The results reveal that roughly 85\% of the
overall energy transfer is primarily attributed to strain-rate
self-amplification ($\{\Pi_{l}^S\}+\{\Pi_{nl}^S\}$), most of which is
local (approximately 70\%), whereas the contribution of vortex
stretching ($\{\Pi_{l}^\omega\} + \{\Pi_{nl}^\omega\}$) accounts for
less than 15\%. The dominant role of the local strain-rate
self-amplification is consistent with the saddle point topology of the
conditional-averaged velocity field from figure~\ref{fig:velocity_3d}
and the staggered arrangement between enstrophy and energy transfer
from figure~\ref{fig:enstrophy}.  Interestingly, the contribution of
$\{\Pi_{l}^\omega\}$ is essentially zero, implying that vortex
stretching acts only from larger-scale strain to slightly
smaller-scale vorticity via the term $\{\Pi_{nl}^\omega\}$. This
phenomenon has been observed in previous studies~\cite{Leung2012,
  Lozano2016, Goto2017, Yao2020}. The contribution of $\Pi_{nl}^C$ is
also negligible, at least for the range of filter widths considered.
This agrees with the space-time average values of $\Pi_{nl}^C$
reported in the literature~\cite{Johnson2020, Johnson2021}.  Although
some sensitivities can be appreciated across the filter widths
(figure~\ref{fig:Pi}a) and Reynolds numbers (figure~\ref{fig:Pi}b),
the conclusions remain robust across the cases investigated.


\section*{Conclusions}
\label{sec:conclusions}

We have studied the three-dimensional structure of the flow
surrounding regions of intense kinetic energy transfer in the inertial
range of isotropic turbulence. To that end, the flow velocity was
low-passed filtered using a Gaussian filter and the characteristic
flow patterns surrounding intense energy transfer events were obtained
by conditionally averaging the flow after proper translation,
rotation, and re-scaling of the frame of reference.

Our findings have revealed that forward intense energy-transfer events
are consistently confined in the high strain-rate region located
between two distinct zones of elevated enstrophy resembling
hairpin-like shapes.  In that region, the local velocity field
associated with the energy transfer exhibits a saddle point topology
characteristic of strain-dominated regions.  Our analysis also
highlights that the primary mechanism driving the cascade in these
regions is strain self-amplification from local and non-local
interactions, which accounts for roughly 85\% of the energy transfer,
whereas vortex stretching remains below 15\%.

The main focus of this work has been on the coherent structure and
physical mechanisms involved in the energy cascade. Nonetheless, our
results can inform decisions about subgrid-scale (SGS) model
developments for LES, especially when the intent is to faithfully
represent the processes involved in the cascade. In such situations,
the coherent structure of the energy cascade must be consistent with
the predominant role of strain-self amplification over vortex
stretching. The high strain-rate region created between two distinct
zones of elevated enstrophy, as reported here, can serve as a
benchmark to evaluate the physical fidelity of SGS models in those
contexts. This level of detail is probably not required for SGS models
that are aimed at capturing only the general characteristics of the
flow, such as mean velocity profiles or mean Reynolds stresses.
Another significant implication for SGS modeling stems from the
observation that, although 70\% of the energy transfer
($\approx\{\Pi_{l}^{S}\}$) is due to local interactions (and thus
resolvable by LES), the remaining 30\% arises from non-local
interactions. These latter cannot be resolved by the LES grid and
therefore need to be modeled, introducing the challenge of the closure
problem.

Our results also suggest that control strategies aiming to enhance or
deplete the energy cascade should target the rate-of-strain tensor
rather than vorticity.  However, this assessment may not be
straightforward, as the rate-of-strain tensor and vorticity are
kinematically linked, meaning that manipulating one could influence
the other. Additionally, any intervention designed to modify the
turbulent flow will inevitably affect the dynamics of both the
rate-of-strain and vorticity, potentially diminishing the
applicability of the coherent structures identified in our study.

Finally, it is worth mentioning that our analysis was based on
instantaneous flow snapshots, which only offer a static glimpse into
the flow dynamics. Time-resolved data are likely necessary to
accurately identify the dynamical relevance of the mechanisms involved
in the energy cascade. Future work will be devoted to establishing
cause-effect relationships between time-resolved coherent structures
and the mechanisms responsible for the transfer of energy among
scales.

\section*{Data availability}
The datasets generated and analyzed during the current study are
available in the Jimenez's Group repository,
\url{https://torroja.dmt.upm.es/turbdata/}

\section*{Acknowledgements}
This work was supported by the National Science Foundation under Grant
No. 2140775.

\section*{Author contributions statement}
A.L.-D. conceived the idea, D.P. conducted the data post-processing
and developed the code, D.P. and A.L.-D. analyzed the results. All
authors reviewed the manuscript.

\section*{Additional information}

\textbf{Competing interests}\\
The authors declare no competing interests.

\bibliography{references}

\section*{Supplementary Information}

\subsection*{Sensitivity to the threshold for $\Pi$-structures}
\label{appendix:thres}

\begin{figure}
  \begin{center}
    \subfloat[]{\includegraphics[width=.26\linewidth]{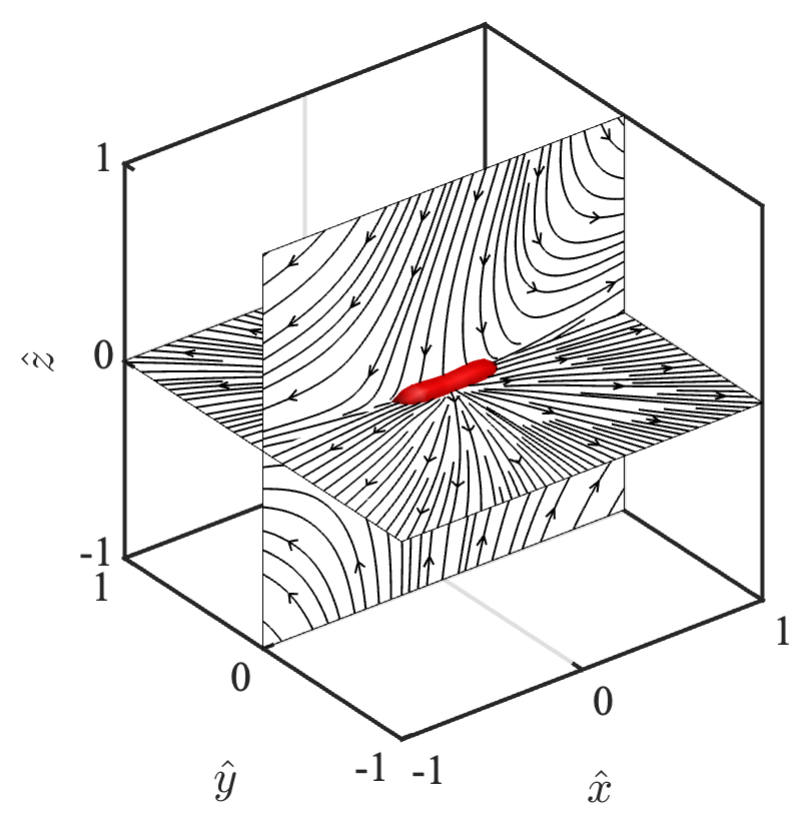}}
    \hspace{0.15cm}
    \subfloat[]{\includegraphics[width=.26\linewidth]{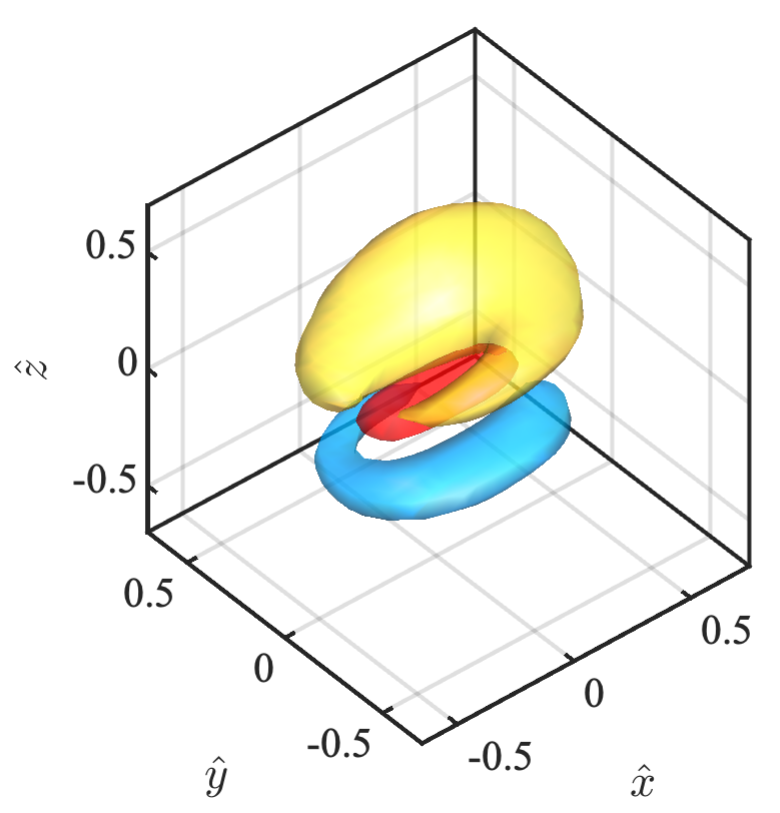}}
    \hspace{0.15cm}
    \subfloat[]{\includegraphics[width=.32\linewidth]{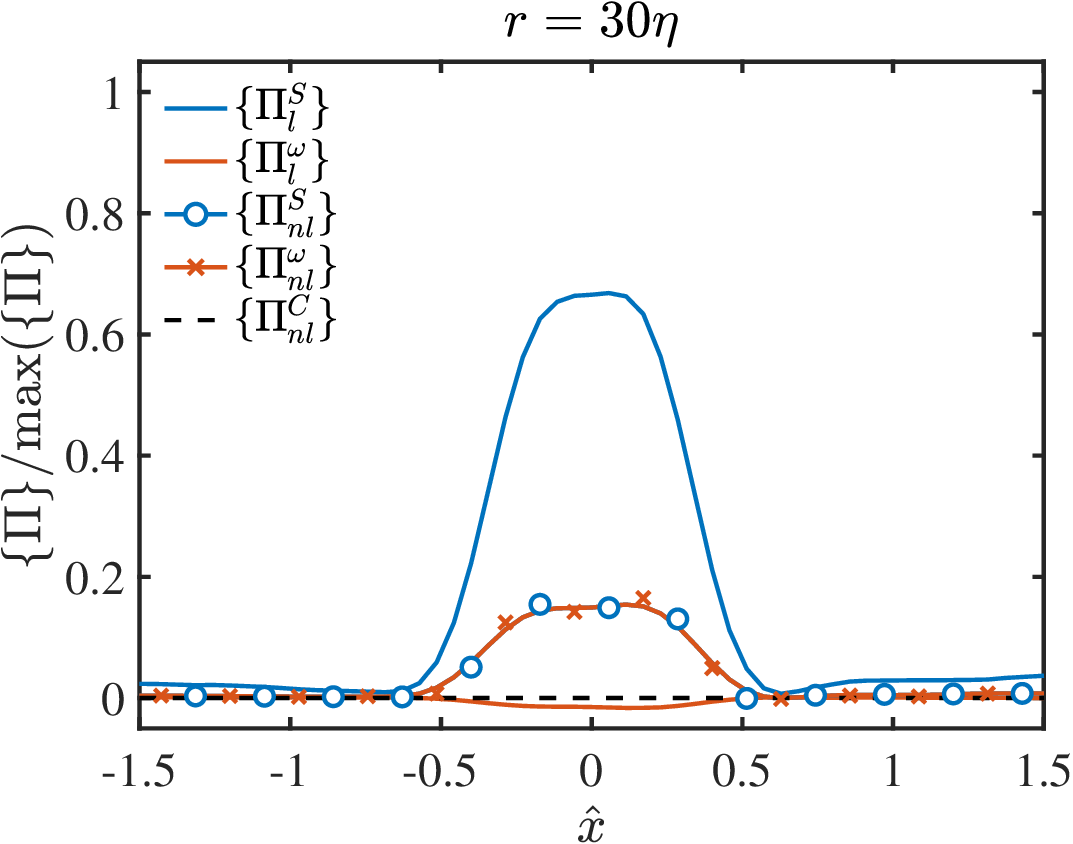}}
  \end{center}
  \begin{center}
    \subfloat[]{\includegraphics[width=.26\linewidth]{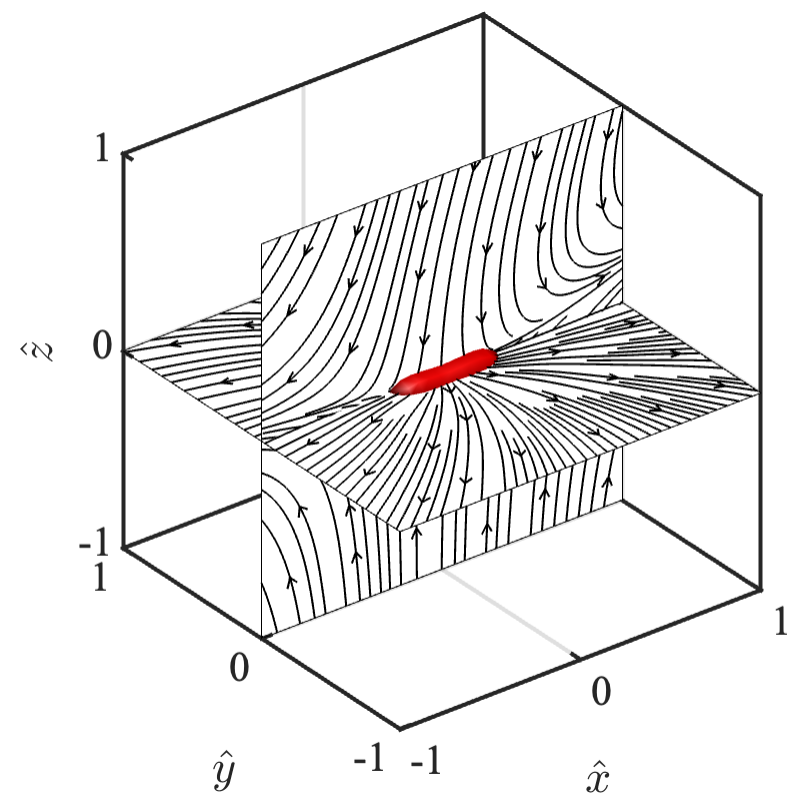}}
    \hspace{0.15cm}
    \subfloat[]{\includegraphics[width=.26\linewidth]{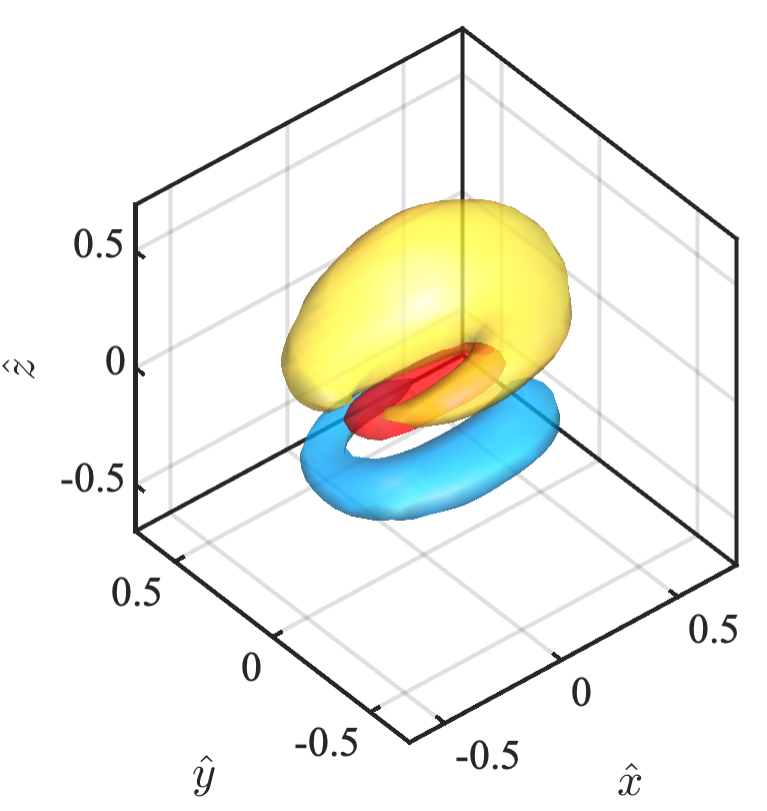}}
    \hspace{0.15cm}
    \subfloat[]{\includegraphics[width=.32\linewidth]{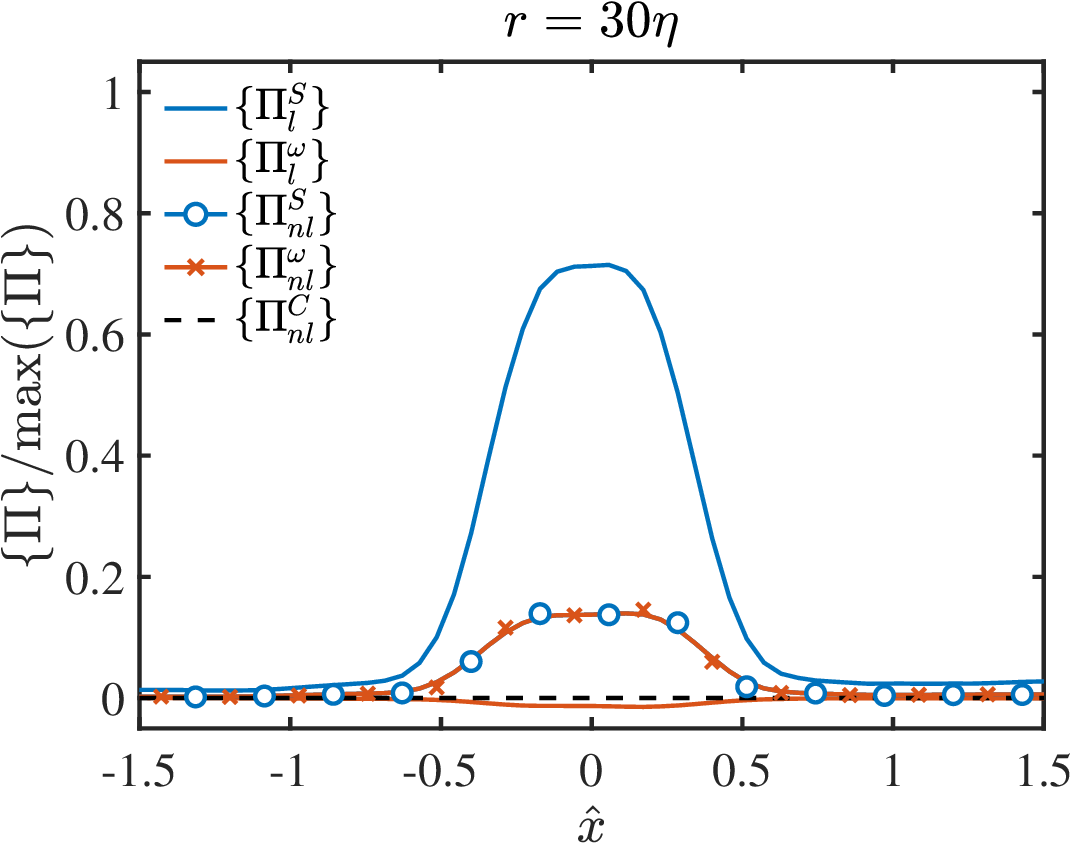}}
  \end{center}
  \caption{Sensitivity to the thresholding parameter to compute
    $\Pi$-structures. Results for conditional-averaged flow fields for
    (top row) $\alpha=1/2$ and (bottom row) $\alpha=2$. The results
    are for HIT3 and $r=30\eta$. \label{fig:appendix:thres}}
\end{figure}

The $\Pi$-structures are defined as regions satisfying $\Pi >
\alpha\Pi_{\text{rms}}$, with the thresholding parameter $\alpha =
1$. We tested the sensitivity of the results for $\alpha = 1/2$ and
$\alpha = 2$. The key results, presented in
figure~\ref{fig:appendix:thres}, show that similar conclusions hold
for both $\alpha = 1/2$ and $\alpha = 2$. This outcome is not
surprising, as $\Pi$-structures are merely used as markers to identify
regions of the flow where most of the energy transfer occurs. These
markers tend to remain in similar locations of the flow across a wide
range of thresholding values.

\subsection*{Statistical uncertainty}
\label{appendix:stats}

\begin{figure}
  \begin{center}
    \subfloat[]{\includegraphics[width=.26\linewidth]{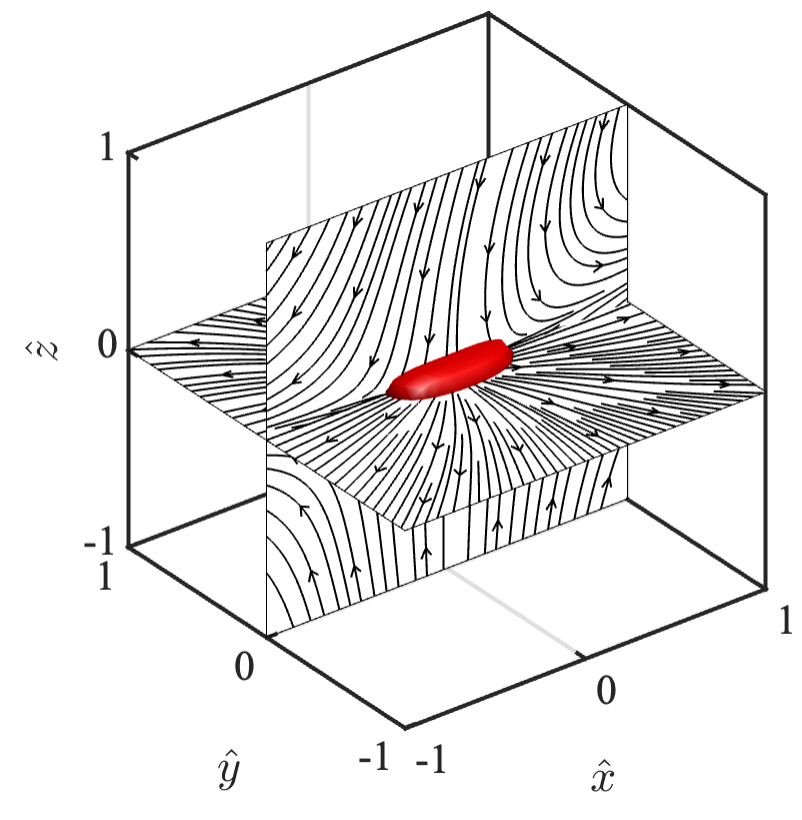}}
    \hspace{0.15cm}
    \subfloat[]{\includegraphics[width=.26\linewidth]{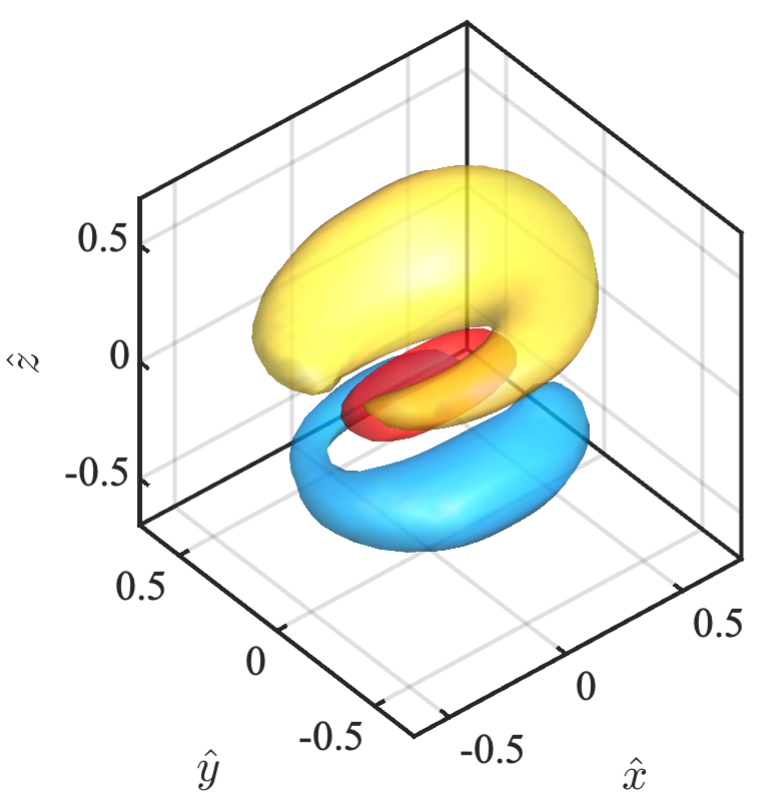}}
    \hspace{0.15cm}
    \subfloat[]{\includegraphics[width=.32\linewidth]{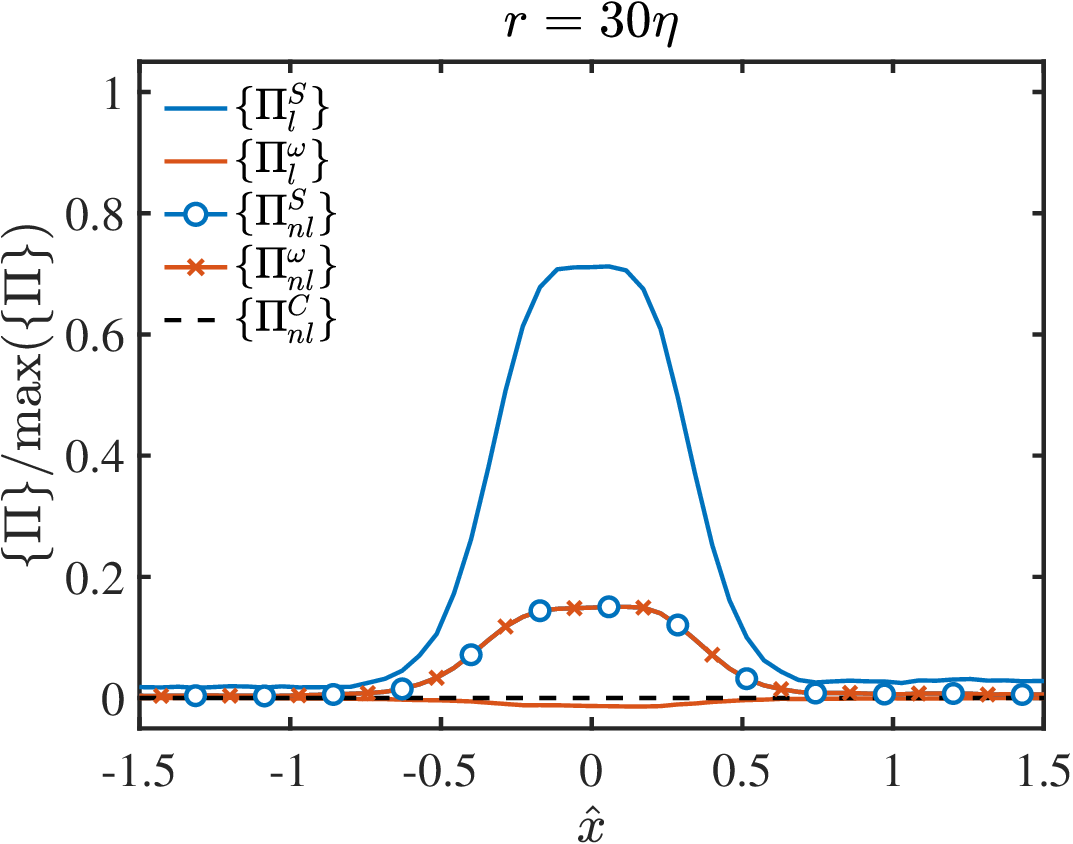}}
  \end{center}
  \caption{Statistical uncertainty. Results for conditional-averaged
    flow fields using one quarter of the total number of samples. The
    results are for HIT3 and
    $r=30\eta$. \label{fig:appendix:uncertainty}}
\end{figure}

The statistical uncertainty is investigated for case HIT3 and
$r=30\eta$. To this end, the number of $\Pi$-structures used to
compute the conditional-averaged fields was decreased to one quarter
of the total number of samples. The key results of the manuscript are
presented in figure~\ref{fig:appendix:uncertainty} for the reduced
dataset. Despite the reduction in the number of samples, the
qualitative structure of the average flow surrounding intense energy
cascade events remains unchanged.

\end{document}